\documentclass[prc,preprint,preprintnumbers,superscriptaddress,nofootinbib]{revtex4}

\usepackage{amsmath,slashed}
\usepackage{graphicx,graphics,xcolor}
\usepackage{dcolumn}
\usepackage[hyperfootnotes=false]{hyperref}
\usepackage{xspace}
\usepackage{enumerate}
\usepackage{varwidth}
\usepackage{comment}
\usepackage{physics}

\usepackage{amssymb}
\usepackage{mathtools}
\usepackage{dsfont}
\usepackage{multirow}

\newcommand{\mpi}{M_\pi}

\newcommand{\mN}{m_N}

\newcommand{\rr}{\mathbf{r}}
\newcommand{\pp}{\mathbf{p}}

\newcommand{\LL}{\mathbf{L}}

\newcommand{\beq}{\begin{equation}}
\newcommand{\eeq}{\end{equation}}

\newcommand{\bsigma}{\boldsymbol{\sigma}}

\newcommand{\mupi}{\mu_\pi}

\newcommand{\inner}[2]{\langle #1 | #2 \rangle}
\newcommand{\avg}[1]{\langle #1 \rangle}

\allowdisplaybreaks[1]

\begin{document}

\preprint{LA-UR-25-28377}

\title{Quantum Monte Carlo calculation of $\delta_{\rm NS}$ in $^{10}$C using an effective field theory approach}

\author{Garrett B. King}
\affiliation{Theoretical Division, Los Alamos National Laboratory, Los Alamos, NM 87545, USA}

\author{Joseph Carlson}
\affiliation{Theoretical Division, Los Alamos National Laboratory, Los Alamos, NM 87545, USA}

\author{Abraham R. Flores}
\affiliation{Department of Physics, Washington University in Saint Louis, Saint Louis, MO 63130, USA}

\author{Stefano Gandolfi}
\affiliation{Theoretical Division, Los Alamos National Laboratory, Los Alamos, NM 87545, USA}

\author{Emanuele Mereghetti}
\affiliation{Theoretical Division, Los Alamos National Laboratory, Los Alamos, NM 87545, USA}

\author{Saori Pastore}
\affiliation{Department of Physics, Washington University in Saint Louis, Saint Louis, MO 63130, USA}
\affiliation{McDonnell Center for the Space Sciences at Washington University in St. Louis, MO 63130, USA}

\author{Maria Piarulli}
\affiliation{Department of Physics, Washington University in Saint Louis, Saint Louis, MO 63130, USA}
\affiliation{McDonnell Center for the Space Sciences at Washington University in St. Louis, MO 63130, USA}

\author{R. B. Wiringa}
\affiliation{Physics Division, Argonne National Laboratory, Argonne, IL 60439, USA}

\begin{abstract}
We compute radiative corrections to the superallowed $\beta$ decay of $^{10}{\rm C}$ in an effective field theory approach using nuclear matrix elements obtained from quantum Monte Carlo calculations. These corrections are an important ingredient in the extraction of the Cabibbo--Kobayashi--Masakawa quark mixing matrix element $V_{ud}$, and the role of this work is to illuminate the uncertainties arising from nuclear structure. Our results provide good agreement with both the traditional extraction of $V_{ud}$, as well as with a more recent evaluation performed using the no-core shell model and a dispersion formalism. The dominant uncertainty in this approach is the presence of two unknown low-energy constants that enter into the relevant nuclear matrix elements. Future determinations of these low-energy constants-- either from QCD or modeling them with two nucleon amplitudes-- would improve the precision of the extraction in this formalism. 
\end{abstract}

\maketitle

\tableofcontents

\section{Introduction}

In the Standard Model (SM) of particle physics, charged-current interactions are mediated by the  Cabibbo--Kobayashi--Maskawa (CKM) quark mixing matrix~\cite{Cabibbo:1963yz,Kobayashi:1973fv}, which is predicted to be unitary.
By independently measuring the individual CKM elements, the unitarity of the CKM matrix can be tested and such tests by now provide powerful probes of physics beyond the Standard Model (BSM) \cite{Czarnecki:2004cw,Towner:2010zz,ParticleDataGroup:2024cfk, Hocker:2001xe,UTfit:2005ras}. 
Unitarity tests of the first row of the CKM matrix, involving the $V_{ud}$, $V_{us}$
and $V_{ub}$ elements,
are among
the most stringent \cite{Gorchtein:2023naa}.
The most precise determination of $V_{ud}$ is currently from superallowed $\beta$ decays \cite{Towner:2010zz,ParticleDataGroup:2024cfk},
with a relative uncertainty of $\delta V_{ud}/V_{ud} = 3 \cdot 10^{-4}$.
Neutron decay  follows closely, with $\delta V_{ud}/V_{ud} = 5 \cdot 10^{-4}$ \cite{Cirigliano:2023fnz}, if the  most precise determinations of the neutron lifetime \cite{Pattie:2017vsj,UCNt:2021pcg,Musedinovic:2024gms} and axial coupling \cite{Markisch:2018ndu} are used. However, discrepancies between beam and bottle measurements of the neutron lifetime \cite{Yue:2013qrc,UCNt:2021pcg}, and between determinations of the nucleon axial charge from different $\beta$ decay correlation coefficients \cite{Beck:2019xye,Markisch:2018ndu} need to be better understood.
$V_{us}$ and the ratio $V_{us}/V_{ud}$ can be extracted from $K \rightarrow \pi \ell \nu$ and $K \rightarrow \mu \nu/\pi \rightarrow \mu \nu$ \cite{FlavourLatticeAveragingGroupFLAG:2024oxs}.  $|V_{ub}|^2 \sim 10^{-5}$ is too small to affect current tests.
Recent theoretical progress in the evaluation of the so-called ``inner radiative corrections'' \cite{Seng:2018qru,Seng:2018yzq,Seng:2020wjq,Czarnecki:2019mwq,Shiells:2020fqp,Hayen:2020cxh}
and of the $K \rightarrow \pi$ form factor at zero momentum
\cite{Carrasco:2016kpy,FermilabLattice:2018zqv,Bazavov:2012cd,RBCUKQCD:2015joy} have led to a
 value of $\Delta_{\rm CKM} = |V_{ud}|^2 + |V_{us}|^2 + |V_{ub}|^2 -1$ which is in tension with the SM at the 
$\sim 2.8\sigma$  level \cite{Seng:2018yzq,Cirigliano:2022yyo,ParticleDataGroup:2024cfk}. This tension, known as the ``Cabibbo angle anomaly'', 
might indicate BSM physics in the 5 to 10 TeV range \cite{Belfatto:2019swo,Grossman:2019bzp,Crivellin:2020lzu,Kirk:2020wdk,Crivellin:2020ebi,Alok:2021ydy,Crivellin:2021bkd,Crivellin:2022rhw,Belfatto:2021jhf,Belfatto:2023tbv,Balaji:2021lpr,Dinh:2023ezl,Branco:2021vhs,Crivellin:2021njn,Gonzalez-Alonso:2016etj,Falkowski:2017pss,Cirigliano:2023nol}.
Since the Cabibbo angle anomaly is mainly driven by theory, however, it is necessary to validate, and, if possible, to further reduce theoretical uncertainties. 

The determination of $V_{ud}$ from the 2020 critical survey of superallowed $\beta$ decays by J.~C.~Hardy and I.~S.~Towner yields 
 \cite{Hardy:2020qwl}
\begin{equation}\label{eq:Vud1}
    |V_{ud}| = 0.97373(27)_{\delta_{\rm NS}} (9)_{\Delta_R^V} (6)_{\delta _R^\prime} (4)_{\delta_{C}} (5)_{\rm exp} = 0.97373(31).
\end{equation}
The largest error comes from the nuclear structure-dependent corrections $\delta_{\rm NS}$. The next largest errors come from the inner radiative corrections, 
$\Delta_R^V$, where the uncertainty is dominated by the nonperturbative contribution to the $W\gamma$ box \cite{Seng:2018qru,Seng:2018yzq,Seng:2020wjq}. 
$\delta_R^\prime$ denotes the ``outer corrections'', which capture contributions from very low-energy photons that do not resolve the structure of the initial and final state nuclei. The last theoretical input is 
$\delta_C$, which describes corrections to the Fermi matrix element arising from isospin breaking in the nuclear potential and wave functions. 
Eq. \eqref{eq:Vud1}  highlights the importance of assigning a reliable error estimate to nuclear theory corrections. This has led to new scrutiny on the nuclear structure dependent corrections  
\cite{Seng:2018qru,Gorchtein:2018fxl,Seng:2022cnq,Gorchtein:2023naa,Cirigliano:2024msg,Cirigliano:2024rfk}, and to the first calculations of $\delta_{\rm NS}$ in light nuclei with $\textit{ab initio}$ nuclear many body methods \cite{Gennari:2024sbn,Cirigliano:2024msg}. In particular, Ref. \cite{Gennari:2024sbn} computed $\delta_{\rm NS}$ for $^{10}$C,
combining the dispersive approach of Ref. 
\cite{Seng:2022cnq} with no-core shell model nuclear structure calculations. 

In this paper, we study nuclear structure-dependent corrections to the decay
of $^{10}$C to $^{10}$B in  the formalism developed in Refs.~\cite{Cirigliano:2023fnz,Cirigliano:2024msg,Cirigliano:2024rfk}, which recast electromagnetic corrections to superallowed decays using a tower of effective field theories (EFTs) 
to capture contributions from photons with different virtualities, which are thus sensitive to different nuclear inputs. In this approach, 
nuclear structure-dependent corrections are induced by photons with three-momentum of order of the typical nucleon momentum in nuclei, $|{\bf q}|  \sim k_F \sim 50 - 100$ MeV. The exchange of these photons gives rise to two- and higher-body corrections to the Fermi operator, which can be organized in a double expansion in
the electromagnetic coupling constant, $\alpha \approx 1/137$, and the chiral expansion parameter $\varepsilon_\chi = Q/\Lambda_\chi$, where $Q$ denotes a scale of order of $k_F$ or the pion mass, $M_\pi$.
The leading contributions to
$\bar\delta_{\rm NS}$ can then be obtained by calculating the matrix elements of two-body currents between the wave functions of the initial and final state nuclei, obtained with chiral interactions that are consistent with the electroweak operator. We follow here the conventions of Ref. \cite{Cirigliano:2024msg},
which denotes with a bar quantities defined in the EFT formalism. The exact mapping between these and the traditional quantities entering the decay rate as given, for example, in Ref. \cite{Hardy:2020qwl}, is spelled out in Table I of Ref.~\cite{Cirigliano:2024msg}.
In this paper we present the first calculation of the nuclear-structure-dependent corrections to the decay of $^{10}$C to $^{10}$B using quantum Monte Carlo (QMC) methods.  

The paper is organized as follows: First, in Section~\ref{sec:eft.ops}, we review the formalism to obtain the operators entering the calculation of $\bar{\delta}_{\rm NS}$. In Section~\ref{sec:qmc}, we provide an overview of the two QMC approaches used in this work; namely, variational (VMC) and Green's function Monte Carlo (GFMC). We then discuss the details of the nuclear Hamiltonians employed in our calculation in Section~\ref{sec:hamiltonian}. Section~\ref{sec:results} contains a discussion of the VMC and GFMC matrix elements, as well as an analysis of the effect of many-body correlations on our results. Taking the QMC results, we discuss the implications for $\bar{\delta}_{\rm NS}$ and $V_{ud}$ in Section~\ref{sec:delta.ns}. Finally, we provide concluding remarks in Section~\ref{sec:conclusions}.

\section{Chiral EFT operators for $\bar\delta_{\rm NS}$}\label{sec:eft.ops}

The calculation of $\bar\delta_{\rm NS}$ in chiral EFT
($\chi$EFT)
can be reduced to the evaluation of matrix elements of two- and three-body transition operators between the wave functions of initial and final states \cite{Cirigliano:2024msg,Cirigliano:2024rfk}. 
The leading contributions to 
$\bar\delta_{\rm NS}$ 
arise at $\mathcal O(\alpha \varepsilon_\chi)$
and $\mathcal O(\alpha \mathcal Q/M_\pi)$, 
where $\mathcal Q$ denotes the reaction $\mathcal Q$-value. The first corrections are induced by transition operators that are independent on the lepton energies, but require insertions of vertices from subleading terms in the chiral Lagrangian. The second type of operators, on the other hand, can be built using only leading-order interactions, but are proportional to the leptonic energies and momenta. We can  write
\begin{equation}
    \bar\delta_{\rm NS} = \delta_{\rm NS}^{(0)} + \overline{\delta_{\rm NS}^{E}},
\end{equation}
where $\delta_{\rm NS}^{(0)}$ is energy-independent, and 
$\overline{\delta_{\rm NS}^{E}}$ results from averaging energy-dependent contributions over the electron phase space.

At $\mathcal O(\alpha \varepsilon_\chi)$ and $\mathcal O(\alpha^2)$, $\delta_{\rm NS}^{(0)}$
can be expressed as
\begin{align}\label{deltaNS0}
   \delta^{(0)}_\text{NS}  &= \frac{2}{g_V(\mupi)M_\text{F}^{(0)}}   \sum_{N = n,p} \bigg[ \alpha  \big( M^\text{mag}_{\text{GT}, N} + M^\text{mag}_{\text{T}, N} +  M^\text{CT}_{\text{GT}, N}    + M_{\text{LS}, N}  \big)      
    + \alpha^2  M^+_{\text{F}, N}\bigg].
\end{align}
Here $M_F^{(0)}= \sqrt{2}$, $\alpha$ is the electromagnetic coupling constant and $g_V$ is the vector coupling constant, evaluated at the renormalization scale $\mu_\pi = M_\pi$,
$g_V(M_\pi)= 1.01494(12)$ \cite{Cirigliano:2023fnz}. 
The error on $g_V$ is dominated by the
nonperturbative contribution
$\overline{\Box}_\text{had}^V (\mu_0)$~\cite{Cirigliano:2023fnz}, which was evaluated with input from Refs.~\cite{Seng:2018qru,Seng:2018yzq,Czarnecki:2019mwq,Shiells:2020fqp,Hayen:2020cxh,Seng:2020wjq,Cirigliano:2022yyo}.
$M^{\mathcal X}_{i, N}$ denote the matrix elements of two-body operators
\begin{align}\label{eq:densities}
M^{\mathcal X}_{i, N} &=\int_0^\infty d r   \, C^{\mathcal X}_{i, N}(r) =    \langle f |  \mathcal V^{\mathcal X}_{i, N} | i \rangle,
\end{align}
where $i = \{\text{F} ,\text{GT}, \text{T}, \text{LS}\}$.
In addition to the matrix elements, we will also show the two-body operator densities $C$, which are implicitly defined by Eq. \eqref{eq:densities}.
With notation that is reminiscent of the neutrinoless double-$\beta$ decay literature \cite{Engel:2016xgb},
these can be expressed in terms of
Gamow-Teller (GT), Fermi (F), Tensor (T) and spin-orbit (LS) components while the isospin structure of the operator is accounted for by splitting them into a $N=p$ and $N=n$ component, denoting the coupling of the photon emitted by the electron or positron to a ``spectator'' proton or neutron. For positron emission, the operators are given by 
\begin{align}
    \mathcal V^{\mathcal X}_{\text{F}, N} &= 
    \sum_{j < k} h_{\text{F},N}^{\mathcal X}(r_{jk}) \mathcal O_{\text{F},N} = 
    \sum_{j < k} h_{\text{F},N}^{\mathcal X}(r_{jk}) \Big[ \tau^{- (j)}  P^{(k)}_N  + (j \leftrightarrow k)  \Big],  \label{eq:OF}\\
    \mathcal V^{\mathcal X}_{\text{GT}, N} &= \sum_{j < k} h_{\text{GT},N}^{\mathcal X}(r_{jk}) \mathcal O_{\text{GT},N} =  \sum_{j < k} h_{\text{GT}, N}^{\mathcal X}(r_{jk}) \, \bsigma^{(j)} \cdot  \bsigma^{(k)} \Big[ \tau^{- (j)}  P^{(k)}_N  + (j \leftrightarrow k)  \Big],
    \label{eq:OGT}
    \\
    \mathcal V^{\mathcal X}_{\text{T}, N} & =  \sum_{j < k} h_{\text{T},N}^{\mathcal X}(r_{jk}) \mathcal O_{\text{T}, N} =   \sum_{j < k} h_{\text{T},N}^{\mathcal X}(r_{jk}) S^{(jk)}({\bf \hat r}) \Big[ \tau^{- (j)}  P^{(k)}_N  + (j \leftrightarrow k)  \Big], \label{eq:OT}
    \\
        \mathcal V^{\mathcal X}_{\text{LS}, N} & =  \sum_{j < k} h_{\text{LS}, N}(r_{j k}) \mathcal O_{\text{LS}, N} =  \sum_{j < k} h_{\text{LS}, N}(r_{j k}) \Big[ \tau^{- (j)} P_N^{(k)}  (\LL_{jk} - \LL^{\text{CM}}_{jk}) \cdot \boldsymbol{\sigma}^{(j)}  + (j \leftrightarrow k) \Big], \label{eq:Oso}
\end{align}
with $\rr_{jk} = \rr_j - \rr_k$,
$r_{jk}= |\rr_{jk}|$, $
S^{(jk)}(\hat \rr)=  3 \hat\rr \cdot \bsigma^{(j)} \, \hat\rr  \cdot \bsigma^{(k)} - \bsigma^{(i)} \cdot \bsigma^{(j)}$,  $
\LL_{jk} = - \frac{i}{2} \rr_{jk}\times \left({\boldsymbol{\nabla}}_j - {\boldsymbol{\nabla}}_k\right)$
and 
$
\LL^{\text{CM}}_{jk} = - \frac{i}{2} \rr_{jk}\times \left({\boldsymbol{\nabla}}_j + {\boldsymbol{\nabla}}_k\right)$
. Notice that when exchanging $j \leftrightarrow k$ $\LL$ remains unchanged, while $\LL^{\text{CM}} \rightarrow - \LL^{\text{CM}}$.
$P_{p,n}$ denote proton and neutron projectors $P_{p,n} = (1\pm \tau_3)/2$.
The dependence on the two-nucleon distance is captured by the radial functions $h(r)$ and the label $\mathcal X \in \{ {\rm mag}, {\rm CT}, + \}$. The operators induced by the nucleon magnetic moments (``mag'') 
and the spin-orbit operators
are predominantly long-range, with radial function $\sim 1/r$. CT denotes short-range, contact contributions, which come with two unknown low-energy constants (LECs), while $M_F^+$ depends logarithmically on the two-nucleon distance.  
The radial functions are given in Appendix \ref{app:potentials}. Notice that the radial function in the coordinate-space form of the spin-orbit term has a different sign compared to Ref.~\cite{Cirigliano:2024msg}\footnote{We thank C.~Y.~Seng for pointing out a sign mistake in the Fourier transform of $\mathcal V_{\rm LS}$ in Ref. \cite{Cirigliano:2024msg}. We further thank W.~Dekens and J.~de~Vries for checking the sign.}. In addition, Ref. \cite{Cirigliano:2024msg}
neglected the $\LL^{\text{CM}}$ term  in
$\mathcal V_{\rm LS}$,
which is proportional to the
 the two-nucleon center of mass momentum. We find that this term is actually non-negligible\footnote{We thank S.~Novario for pointing out that the contribution of $\LL^{\text{CM}}$ does not vanish in many-body systems.}.

In the EFT formalism $\bar\delta_{\rm NS}$ also receives an energy-dependent correction at $\mathcal O(\alpha \mathcal Q/M_\pi)$.
This correction is given by
\begin{align}\label{deltaNSE}
\overline{\delta^E_\text{NS}} &= \mp \alpha \frac{2}{g_V(\mupi)M_\text{F}^{(0)}} R_A E_0
\sum_{N=n,p} \bigg[
\tilde{f}_E  M^E_{\text{F}, N}  + 
 M^{E\pi}_{\text{GT}, N}+ M^{E\pi}_{\text{T}, N} +
\tilde f_{m_e}^\pi
 \left( M^{m_e\pi}_{\text{GT}, N}+ M^{m_e\pi}_{\text{T}, N}\right)\bigg],
\end{align}
where the upper sign is for $\beta^+$, the lower sign for $\beta^-$ decays.
Here $R_A = 1.2 A^{1/3}$ fm is introduced to make the matrix elements dimensionless.
$E_0$ is the electron endpoint energy, 
\begin{equation}
    E_0 = \mathcal Q_{\rm EC} - m_e,
\end{equation}
in terms of the electron capture 
$\mathcal Q_{\rm EC}$ value given in Ref.~\cite{Hardy:2020qwl}, $\mathcal Q_{\rm EC} = 1907.994 \pm 0.067$ KeV.
The factors 
$\tilde f_E$
and $\tilde f_{m_e}^\pi$
arise from the phase-space average, and are given by
\begin{align}
 \tilde{f}_E &= \frac{1}{E_0}   \left(\frac{4}{3}\langle E_e \rangle + \frac{1}{6} E_0 + \frac{1}{2}\left\langle \frac{m^2_e}{ E_e} \right\rangle \right),\notag\\
 \tilde{f}_{m_e}^\pi &=    \frac{1}{E_0}\left\langle \frac{m^2_e}{E_e} \right \rangle, 
\end{align}
where the phase-space average is defined in Eq. \eqref{eq:PSaverage2}.
For $^{10}$C $\rightarrow$ $^{10}$B, we find
\begin{equation}
   \tilde f_E = 1.105, \qquad \tilde f_{m_e}^\pi = 0.23.
\end{equation}
In the following, we will calculate the matrix elements entering Eqs. \eqref{deltaNS0} and \eqref{deltaNSE} for the $^{10}{\rm C}(0^+) \rightarrow ^{10}{\rm B^*}(0^+;1)$ transition, using the VMC and GFMC methods.

The derivation of the EFT contributions to $\bar\delta_{\rm NS}$  is very similar to the approach used for
the calculation of $\delta_{\rm NS,\, B}$, one component of the nuclear-structure-dependent corrections included  in Refs.~\cite{Hardy:2020qwl,Towner:1992xm}. The main difference is that in the approach of Ref.~\cite{Hardy:2020qwl,Towner:1992xm} the short-range behavior of the electroweak transition operator is regulated by the nucleon axial and magnetic form factors, while the chiral EFT operators
acquire a short-range component which contains unknown LECs that must be extracted from data or calculated from QCD \cite{Cirigliano:2024msg,Cirigliano:2024rfk}.
The LECs parameterize all two-nucleon physics at $|{\bf q}| \gg k_F$, and thus they capture at least part of the $g_A$ and $g_M$  ``quenching''  
included in $\delta_{\rm NS,\, A}$. Since the calculation of
$\bar\delta_{\rm NS}$ in one isotope does not provide enough information to extract the LECs, the comparison of these components cannot at present be made more precise.

\section{Quantum Monte Carlo methods}\label{sec:qmc}

To obtain the many-body wave functions necessary to compute matrix elements of the EFT transition operators, we use quantum Monte Carlo (QMC) approaches. QMC approaches generally consist of a suite of stochastic methods to solve the Schr\"{o}dinger equation for strongly-correlated many-body systems. Their application to the structure of finite nuclei has been reviewed extensively~\cite{Carlson:2014vla,Gandolfi:2020pbj}, and more recently their application to the study of electroweak processes was reviewed in Ref.~\cite{King:2024zbv}. Here, we will overview the salient features of the approaches, and direct interested readers to the aforementioned review articles for more details. 

The QMC approach for finite nuclei involves two steps, the first of which is variational Monte Carlo (VMC). Ultimately, we would like to obtain a state $\Psi(J^\pi;T,T_z)$, with specific spin-parity $J^\pi$, isospin $T$, and isospin projection $T_z$ quantum numbers. To achieve this, we make an ansatz for a trial wave function $\Psi_T$ of the form~\cite{Pudliner:1997ck},
\begin{equation}
\ket{\Psi_T} = \mathcal{S}\prod_{i<j}\left[ 1 + U_{ij} + \sum_{i<j\neq k} \widetilde{U}^{\rm TNI}_{ijk} \right] \ket{\Psi_J}\, ,\label{eq:psi.t}
\end{equation}
where $\mathcal{S}$ is the symmetrization operator, $U_{ij}$ is a two-body correlation operator, $\widetilde{U}^{\rm TNI}_{ijk}$ is a three-body correlation operator, and $\Psi_J$ is a Jastrow-like wave function. Encoded in $\Psi_J$ are the global properties of the system; namely, antisymmetry and the quantum numbers of the state. The correlation operators reflect the influence of the nuclear interaction at short-range, and embedded within them are variational parameters that are optimized by minimizing the energy expectation value,
\begin{equation}
E_T = \frac{\mel{\Psi_T}{H}{\Psi_T}}{\inner{\Psi_T}{\Psi_T}}\, ,
\end{equation}
for the nuclear Hamiltonian $H$. This will give a state with an energy that provides an upper bound to the true ground state energy of the system, $E_0$. 

One can improve upon the best variational state $\Psi_V$ with the Green's Function Monte Carlo (GFMC) method. The method leverages the fact that the real time Schr\"{o}dinger Equation may be recast as a diffusion equation in imaginary time $\tau$. Noting that one may expand any state in a complete orthonormal basis, we could imagine $\Psi_V$ as being a linear combination of the true eigenstates $\psi_i$ of $H$,
\begin{equation}
\ket{\Psi_V} = \sum_{i=0}^{\infty} c_i\ket{\psi_i}\, .
\end{equation}
While we do not know the states $\psi_i$, we can project out the true ground state by propagation in $\tau$,
\begin{equation}
\lim_{\tau\to\infty} e^{-(H-E_0)\tau}\ket{\Psi_V} \propto c_0\psi_0\, .
\end{equation}
In practice, propagation is performed in $n$ small steps in imaginary time $\Delta\tau$,
\begin{equation}
\ket{\Psi(\tau)} = \left[ e^{-(H-E_0)\Delta\tau}\right]^n\ket{\Psi_V}\, ,
\end{equation}
until spurious contamination is removed from the wave function and convergence is reached. While one would like to compute diagonal matrix elements of the form,
\begin{equation}
    \avg{\mathcal{O}(\tau)} = \frac{\mel{\Psi(\tau)}{\mathcal{O}}{\Psi(\tau)}}{\inner{\Psi(\tau)}{\Psi(\tau)}}\, ,
\end{equation}
this would require an imaginary time propagation for each matrix element under study, and would be computationally prohibitive. To circumvent this limitation and perform an analysis of each contribution to the EFT transition operator, we compute mixed estimates. If we take $\Psi(\tau) = \Psi_V + \delta \Psi$, where $\delta \Psi$ is a small correction, then we may write~\cite{Carlson:2014vla,Gandolfi:2020pbj},
\begin{equation}
\avg{\mathcal{O}(\tau)} \approx 2 \frac{\mel{\Psi(\tau)}{\mathcal{O}}{\Psi_V}}{\inner{\Psi(\tau)}{\Psi_V}} - \frac{\mel{\Psi_V}{\mathcal{O}}{\Psi_V}}{\inner{\Psi_V}{\Psi_V}}\, . \label{eq:mixed.diag}
\end{equation}
Note that Eq.~\eqref{eq:mixed.diag} is exact for operators that commute with the Hamiltonian, but is approximate for all other operators. 

While the above applies for diagonal matrix elements, for $\beta$ decay, one must propagate the matrix element,
\begin{equation}
\avg{\mathcal{O}(\tau)}_{i\to f } = \frac{\mel{\Psi_f(\tau)}{\mathcal{O}}{\Psi_i(\tau)}}{\sqrt{\inner{\Psi_f(\tau)}{\Psi_f(\tau)}}\sqrt{\inner{\Psi_i(\tau)}{\Psi_i(\tau)}}}\, ,
\end{equation}
where $\Psi_{i(f)}(\tau)$ is the initial (final) state involved in the transition at imaginary time $\tau$. The mixed estimate was first generalized for off-diagonal transitions in Ref.~\cite{Pervin:2007sc}, and the resultant expression that we use in this work is,
\begin{eqnarray}\nonumber
\avg{\mathcal{O}(\tau)}_{i \to f} &\approx& \sqrt{\frac{\inner{\Psi_f}{\Psi_f}}{\inner{\Psi_i}{\Psi_i}}}\frac{\mel{\Psi_f(\tau)}{\mathcal{O}}{\Psi_i}}{\inner{\Psi_f(\tau)}{\Psi_f}} + \sqrt{\frac{\inner{\Psi_i}{\Psi_i}}{\inner{\Psi_f}{\Psi_f}}}\frac{\mel{\Psi_i(\tau)}{\mathcal{O}^{\dagger}}{\Psi_f}}{\inner{\Psi_i(\tau)}{\Psi_i}}\\ 
&&- \frac{\mel{\Psi_f}{\mathcal{O}}{\Psi_i}}{\sqrt{\inner{\Psi_f}{\Psi_f}}\sqrt{\inner{\Psi_i}{\Psi_i}}}\, . \label{eq:mixed.off.diag}
\end{eqnarray}
The quantity in Eq.~\eqref{eq:mixed.off.diag} is what is propagated in imaginary time, and then averaged once convergence is reached.

\section{Nuclear Hamiltonians}\label{sec:hamiltonian}

Generically, the nuclear Hamiltonian has the form,
\begin{equation}
 H = \sum_{i} T_i + {\sum_{i<j}} v_{ij} + \sum_{i<j<k}
V_{ijk} + \ldots \ ,
\end{equation}
where $T_i$ is the single-nucleon nonrelativistic kinetic energy and $v_{ij}$ and $V_{ijk}$ are two- and three-nucleon potentials, respectively. In this section, we discuss two different potential models: the phenomenological Argonne $v_{18}$ (AV18)~\cite{Wiringa:1994wb} two-body interaction with either the Urbana X (UX)~\cite{Wiringa:2014} or Illinois-7 (IL7)~\cite{Carlson:2014vla} three-body interactions -- referred to collectively as the AV18+UX and AV18+IL7 models -- and two versions of the Norfolk two- and three-nucleon interaction (NV2+3)~\cite{Piarulli:2014bda,Piarulli:2016vel,Piarulli:2017dwd,Baroni:2018fdn}. 

The AV18 is a purely phenomenological model of the nuclear interaction. Included in the potential are the dominant one-pion exchange (OPE) contributions, as well as intermediate-range contributions that approximate two-pion exchange (TPE) with radial dependencies that assume this exchange is dominated by the tensor Yukawa contribution. Finally, there is a short-range potential whose shape is given by a Woods-Saxon form~\cite{Wiringa:1994wb}. All together, there are 18 different operator structures and 42 parameters in the potential. These parameters were constrained with 4301 np and pp scattering data from the Nijmegen partial wave analysis~\cite{Stoks:1993tb} with a $\chi^2$/datum of ${\sim}1.1$. The UX is a phenomenological model of the three-nucleon force that hybridizes two other models; namely, the IL7 and Urbana IX (UIX) models~\cite{Carlson:2014vla}. The UX supplements the long-range two-pion P-wave and central S-wave repulsion of the UIX with a two-pion S-wave term, taking the strengths of all three of these terms from the IL7 parametrization. The IL7 contains three-pion ring diagrams involving one or two intermediate $\Delta$'s, in addition to the terms of the UX model, and is fit to ground state and low-lying excitation energies of $A\le 10$ nuclei. 

The NV2+3 is a semi-phenomenological model based on a chiral effective field theory approach that retains nucleons, pions, and $\Delta$-isobars as degrees of freedom. Because of the natural separation between the typical momentum scale of nucleons inside the nucleons, which is on the order of the pion mass $M_{\pi}$ and $N-{\rm to}-\Delta$ mass splitting, compared to the scale of heavier mesons $\Lambda_{\chi}$, one can integrate out the heavy degrees of freedom. Further, the approach allows one to order contributions to the force in powers of $M_{\pi}/\Lambda_{\chi}$, which would in principle give rise to a systematic expansion; however, it is worth noting that the precise details of the power counting remain an open question~\cite{Hammer:2019poc}, and thus one must choose a scheme to organize the terms. The NV2 potential was derived by first developing a minimally nonlocal two-body force by means of Fierz transformations up to N$^3$LO ({\it i.e.}, to $\mathcal{O}(\varepsilon_\chi^3)$)~\cite{Piarulli:2014bda} in the power counting prescription of Weinberg~\cite{Weinberg:1991um}. Because the use of local potentials are more efficacious for QMC approaches~\cite{Carlson:2014vla}, only those local terms at N$^3$LO necessary to provide a good description of nucleon-nucleon scattering were retained to form the semi-phenomenological NV2 model~\cite{Piarulli:2016vel}.
This is supplemented by an N$^3$LO ($\mathcal{O}((\varepsilon_\chi^3)$) three-body force, the NV3, based on terms first derived by van Kolck {\it et al.}~\cite{vanKolck:1994yi} and Epelbaum {\it et al.}~\cite{Epelbaum:2002vt}. In the $\Delta$-full picture, the diagram involving the excitation of an intermediate $\Delta$-- corresponding to the Fujita-Miyazawa (FM) term~\cite{Fujita:1957zz} is promoted to NLO ($\mathcal{O}(\varepsilon_\chi)$) in the chiral expansion~\cite{Piarulli:2019cqu}. In total, NV2+3 has 26 unknown low-energy constants (LECs) that parameterize the underlying QCD of the contact terms in the short-range two-body potential, plus two additional unknown three-body LECs. Various fitting schemes were used to obtain a suite of NV2+3 interactions; namely, choices for which two-nucleon data to fit, how to regularize singularities in the long-range terms of the potential, and how to fit the three-body force. In this work, we focus on two interactions -- the NV2+3-Ia and NV2+3-Ia$^{\star}$ -- which fit the NV2-Ia two-body force with the same data and regularization scheme, but differ in how the three-body force was fit. In particular, the two-body force is fit to ${\sim}$2700 np scattering data from the Granada database~\cite{Perez:2013jpa,Perez:2013oba,Perez:2014yla}, and like the AV18, has a $\chi^2/$datum of ${\sim}1.1$. The NV3 fits the three-body force to the trinucleon binding energies and the $n-d$ doublet scattering length~\cite{Piarulli:2017dwd}, while the NV3$^{\star}$ was fit to the trinucleon binding energies and the tritium Gamow-Teller $\beta$ decay matrix element~\cite{Baroni:2018fdn}. The NV2-Ia combined with the former fitting scheme is denoted as the NV2+3-Ia interaction, while the combination with the latter fitting scheme is called NV2+3-Ia$^{\star}$~\cite{Piarulli:2016vel}. 

Having two models of the two-nucleon interaction, as well as different three-body forces, allows us to investigate how different modeling choices impact the predictions of $\bar\delta_{\rm NS}$ and $V_{ud}$. Namely, we can investigate how different approaches to regularizing short-range singularities in the interaction and different three-nucleon forces impact our calculation. The AV18 is a relatively hard interaction, with the central potential taking values up to ${\sim}2$ GeV~\cite{Carlson:2014vla}. The NV2-Ia, instead, has a softer core and the regularization of short-range singularities corresponds to a cutoff of ${\sim}500$ MeV in momentum space~\cite{Piarulli:2016vel}. The different three-body fitting schemes can also have an impact on the prediction of static observables in light nuclei, as the AV18+IL7 models and the NV2+3-Ia produce accurate energy spectra of light nuclei with $A\leq 12$ compared to their counterparts~\cite{Carlson:2014vla,Piarulli:2017dwd}. With these four models, we are able to make an {\it ad hoc} uncertainty estimate for the predicted quantities; however, this certainly does not constitute a robust quantification of uncertainties arising from the nuclear interaction. Although great effort has been spent in the development of nuclear interactions with robustly quantified uncertainties~\cite{Bub:2024gyz,Somasundaram:2023sup,Thim:2023fnl,Svensson:2023twt}, and emulators have been designed for various nuclear observables~\cite{Konig:2019adq,Wesolowski:2021cni,Odell:2023cun,Becker:2023dqe,Somasundaram:2024zse,Armstrong:2025tza}, such an analysis would be beyond the scope of the present study.

\section{Results}\label{sec:results}

\subsection{VMC operator densities}\label{sec:vmc}

\begin{table}
\begin{tabular}{||c| c | c | c | c |  c | c | c || c|c||}
\hline
Model & Method & $M^{\text{mag}}_{\text{GT}, p}$ & $M^{\text{mag}}_{\text{GT}, n}$ & $M^{\text{mag}}_{\text{T}}$ & $M_{\text{LS}, p}$& $M^{\text{CT}}_{\text{GT}, p}$ & $M^{\text{CT}}_{\text{GT}, n}$  & $M_{\text{F}, p}^+$  & $M_{\text{F},p}^{E}$\\ 
\hline 
NV2+3-Ia  & VMC & $-0.461$ & $0.064$ & $0.007$ & $-0.006$ & $0.050$ & 
$0.018$ & $-3.32$ & $-4.915$
\\
&GFMC &$-0.515$ & $0.082$ & $0.002$ & $-0.005$ & $0.051$ & 
$0.025$ & $-2.83$ & $-4.74$ \\
\hline 
NV2+3-Ia$*$ & VMC & $-0.469$ & $0.057$ &   $0.005$ & $-0.004$ & $0.045$ & $0.016$ &  $-4.13$ & $-5.201$ \\
&GFMC &$-0.486$ & $0.074$ &$0.000$ & $-0.006$ & $0.044$ & $0.021$ & $-4.13$ & $-5.235$ \\
\hline
\hline 
AV18+UX & VMC & $-0.496$ & $0.053$ & $0.007$ & $-0.005$ & $0.052$ & $0.014$ & $-3.54$ & $-4.955$\\
&GFMC & $-0.455$ & $0.060$ & $0.000$ & $-0.004$ & $0.032$ & $0.015$ &$-5.43$& $-5.692$  \\ 
\hline
AV18+IL7 & VMC & $-0.503$ & $0.053$ & $0.008$ & $-0.005$ & $0.045$ & $0.015$ & $-3.48$ & $-4.938$ \\
&GFMC &$-0.492$ &$0.069$ & $0.002$ & $-0.004$ & $0.043$ & $0.020$ & $-3.96$ & $-5.160$ \\ \hline
\end{tabular}
\caption{Nuclear matrix elements entering $\bar\delta_{\rm NS}$.
The GT, T and LS matrix elements 
contribute to the energy-independent component, $\delta_{\rm NS}^{(0)}$, at $\mathcal O(\alpha)$.
$M_{\text{F},p}^+$ contributes to $\delta_{\rm NS}^{(0)}$ at $\mathcal O(\alpha^2)$. 
$M_{\text{F},p}^{E}$ gives the largest contribution to the energy-dependent component, 
$\overline{\delta^E_{\rm NS}}$. Pion-range contributions to $\overline{\delta^E_{\rm NS}}$ are small, and are given in Appendix \ref{app:a3}.}\label{TabI}
\end{table}

\begin{figure}
    \centering
    \includegraphics[width=\linewidth]{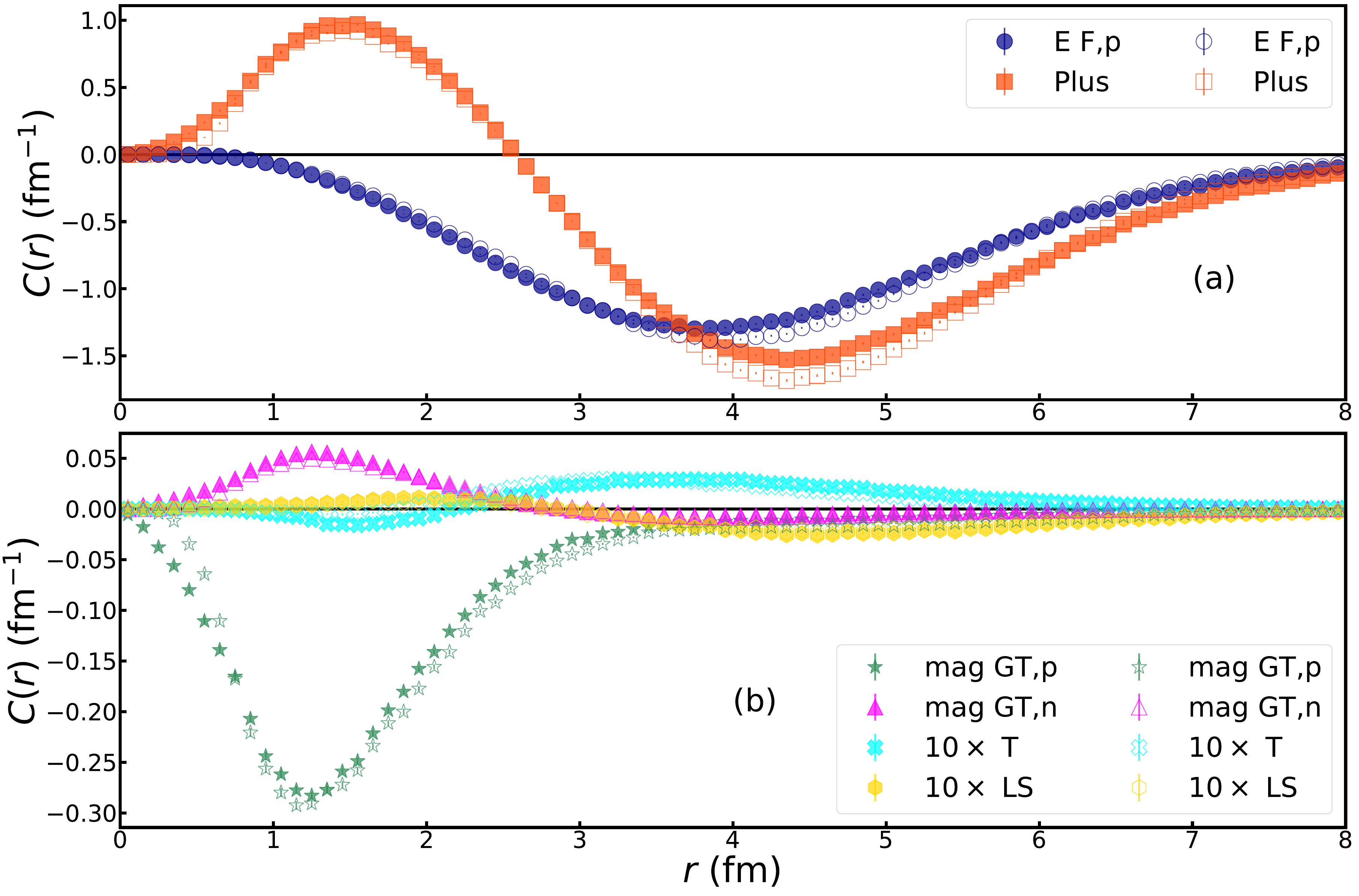}
    \caption{Comparison of the matrix elements obtained with the Ia interaction  (filled symbols) vs 
AV18+UX (empty symbols) computed using VMC. 
    }
    \label{fig:av18}
\end{figure}

\begin{figure}
    \centering
    \includegraphics[width=\linewidth]{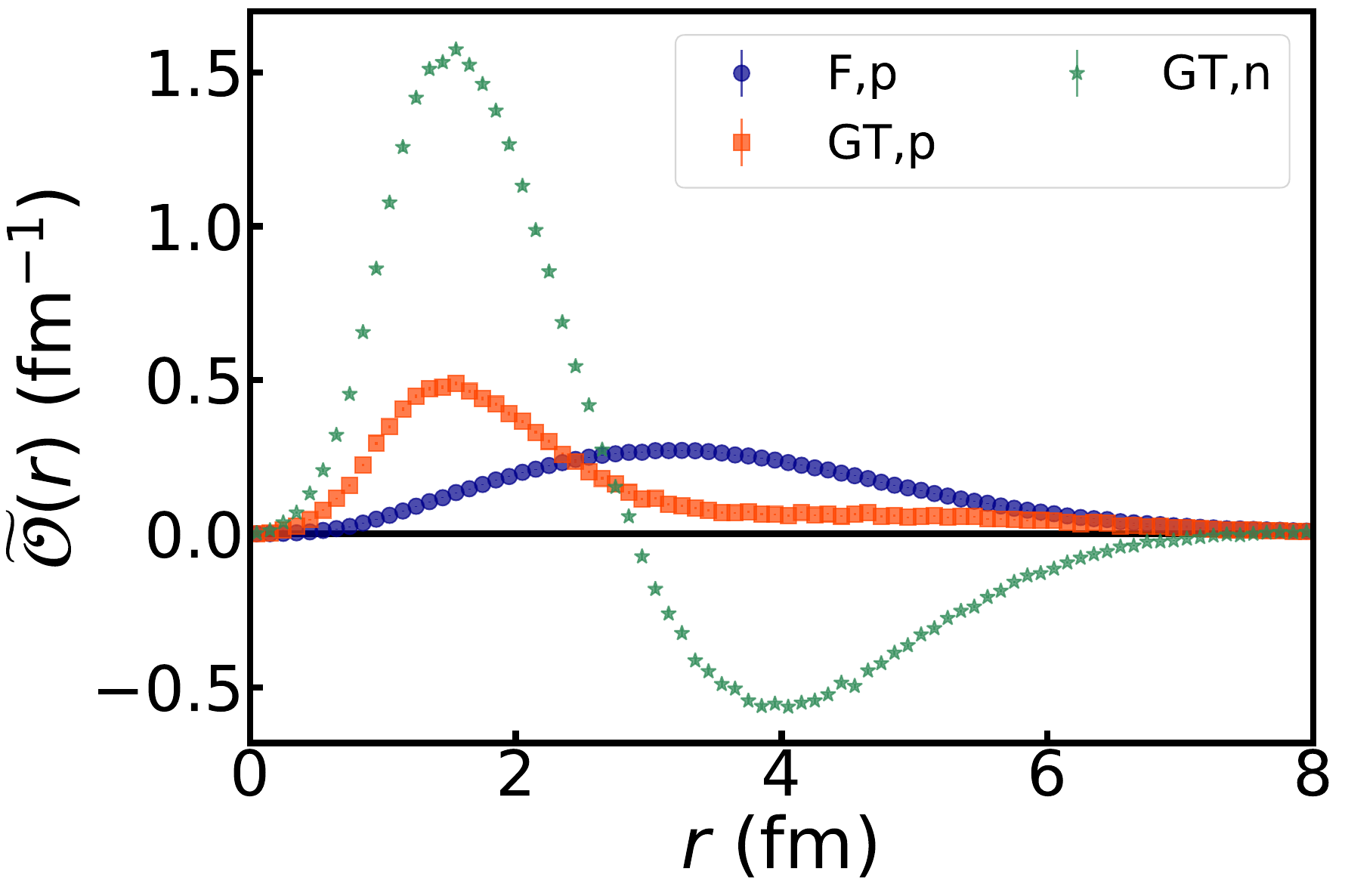}
    \caption{Normalized operator densities according to Eq.~\eqref{eq:o.tilde} computed with VMC using the NV2+3-Ia interaction.}
    \label{fig:operators}
\end{figure}

\begin{table}
\begin{tabular}{|| c | c | c | c ||}
\hline
Method  &$\mathcal{M}_{\text{F}}$ & $\mathcal{M}_{\text{GT}, p}$ & $\mathcal{M}_{\text{GT}, n}$ \\ 
\hline 
VMC  &$7.07$ &$-2.33$ &$0.29$  \\
\hline
\end{tabular}
\caption{Matrix elements for the main operator structures contribution to $\bar\delta_{\rm NS}$ computed with the NV2+3-Ia model.}\label{tab:op.mels}
\end{table}

\begin{figure}
    \centering
    \includegraphics[width=\linewidth]{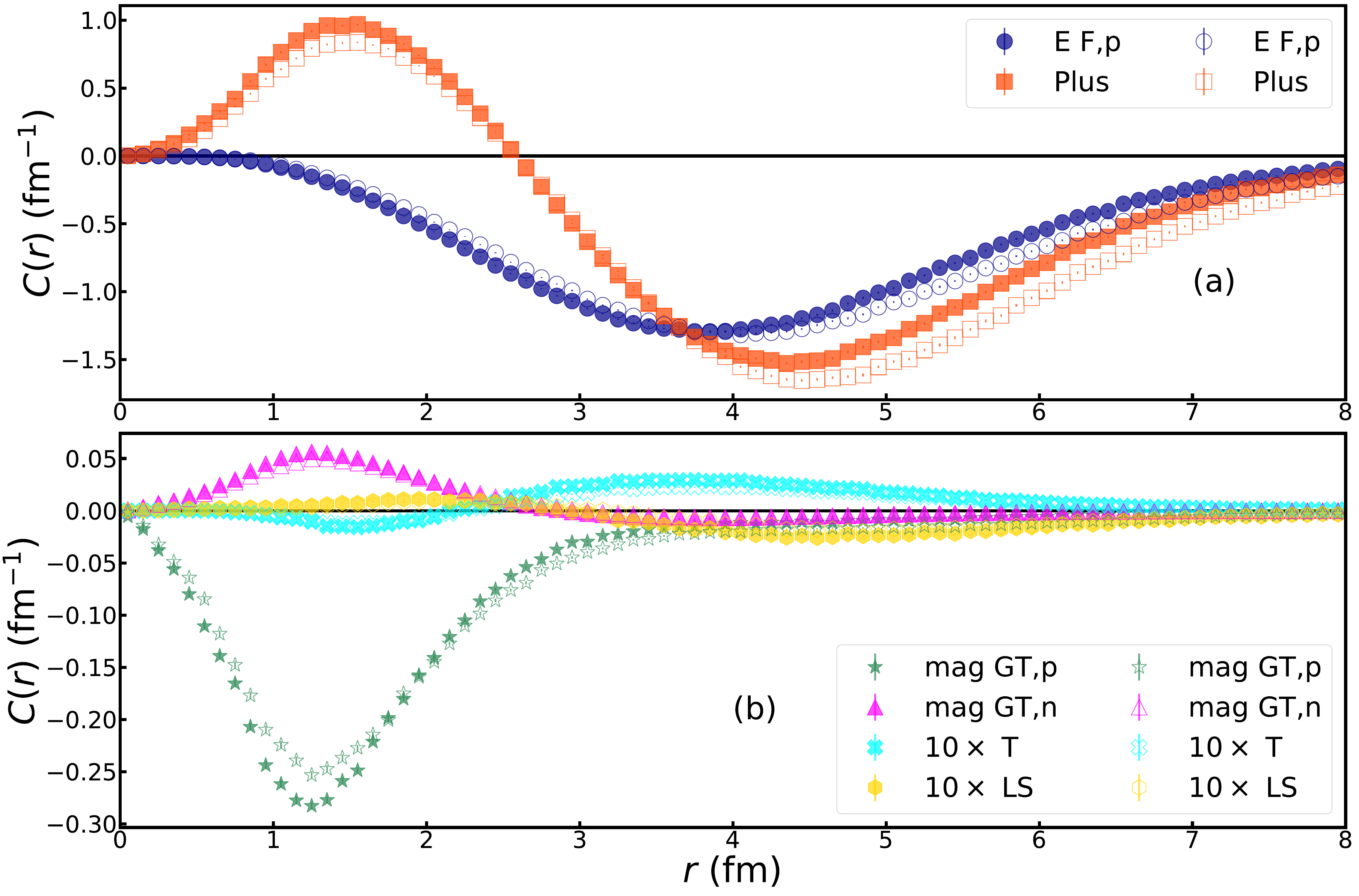}
    \caption{Comparison of the matrix elements obtained with the Ia interaction  (filled symbols) vs 
Ia* (empty symbols) computed using VMC. 
    }
    \label{fig:chi}
\end{figure}

We present the dominant matrix elements computed with both VMC and GFMC in Table~\ref{TabI}.
Matrix elements of pion-range operators, which, as already observed in the case of $^{14}$O \cite{Cirigliano:2024msg}, turn out to be small, are shown in Table
~\ref{TabII} in Appendix \ref{app:a3}. In order to better analyze the model dependence present in our results, we also computed transition densities $C(r)$ according to Eq.~\eqref{eq:densities}. In the following section, we discuss the features of these densities and their impact on the resultant matrix elements. 

First, we compare one chiral model and one phenomenological model-- the NV2+3-Ia and AV18+UX, respectively-- in Figure~\ref{fig:av18}. Panel (a) displays the densities for the spin-independent matrix elements $M^E_{\text{F},p}$ and $M^+_{\text{F},p}$, while panel (b) shows the spin-dependent matrix elements contributing to the energy-independent part of $\bar\delta_{\rm NS}$.  We plot the AV18+UX results with open symbols, and use the filled symbols to represent the results obtained when using the NV2+3-Ia interaction. We see that there is a depletion of short-range strength when using the AV18+UX, which may be due to the fact that the two-body interaction has a hard core when compared with the chiral potential and depletes the relative wave function of pairs at short distances~\cite{Cruz-Torres:2019fum}. This could, in turn, result in some of the strength being shifted out to large interparticle spacings; however, because the three-nucleon interactions are also different, we note that this could also have some interplay with the two-nucleon interaction and participate in the shifting of strength to larger separations. 

We note the presence of nodes in several transition densities. To better interpret the densities and the origin of these nodes, we include plots of the operator densities without radial functions. The operator density is computed in VMC as 
\begin{equation}
    \mathcal{O}_{\alpha}(r) = \frac{1}{4\pi r^2}\sum_{i<j} \mel{\Psi}{\mathcal{O}_{\alpha}\delta(r-|\rr_i-\rr_j|)}{\Psi}\, ,
\end{equation}
with $\mathcal O_\alpha$ defined by Eqs. \eqref{eq:OF}
-- \eqref{eq:Oso}. Fig.~\ref{fig:operators} shows the NV2+3-Ia operator densities scaled by by their integral; {\it i.e.}, we plot,
\begin{equation}
    \widetilde{\mathcal{O}}_{\alpha}(r) = \frac{\mathcal{O}_{\alpha}(r)}{\mathcal{M}_{\alpha}}\, , \label{eq:o.tilde}
\end{equation}
where $\mathcal{M}_{\alpha}$ is simply,
\begin{equation}
\mathcal{M}_{\alpha} = \int_0^{\infty} dr \mathcal{O}(r)\, ,
\end{equation}
so that we may visualize all densities in the same plot. The matrix elements $\mathcal{M}_{\alpha}$ are reported in Table \ref{tab:op.mels} for the NV2+3-Ia optimal VMC wave function.

To validate the implementation of the Fermi matrix element, we note that we anticipate $\mathcal{M}_{\rm F} = \sqrt{2}(Z-1) \approx 7.07$, where $Z = 6$ is the number of protons in the initial state. Indeed, we see in Table~\ref{tab:op.mels} that this relation holds. 

The situation is more complex for the spin-dependent matrix elements. In particular, the operator $\mathcal{O}_{\text{GT},n}$ displays a node around $r\approx 3$ fm in Fig.~\ref{fig:operators}, which is consistent with the transition densities in Figs.~\ref{fig:av18} and~\ref{fig:chi}. This is because the sign of the matrix element depends on whether the pair is in a relative $S=0$ or $S=1$ pair. $\mathcal{O}_{\text{GT},p}(r)$ and $\mathcal{O}_{\text{GT},n}(r)$ both have negative values at short distances. For the former, this can be understood on the basis of short-range $pp$ pairs tending to form $S=0$ states inside the nucleus~\cite{Forest:1996kp}. In the case of the latter, both $S=1$ and $S=0$ $np$ pairs can form at short distances. However, as the magnitude of the $S=0$ contribution is three times greater than for $S=1$, the ratio of short-range pairs is such that the overall sign is negative. At longer ranges, the competition of $S=0$ and $S=1$ pairs will influence the distribution as the $S=1$ $np$ pairs become sufficient to overtake the $S=0$ $np$ pairs, driving the sign change in $\mathcal{O}_{\text{GT},n}(r)$. Because the formation of pairs is sensitive to the details of the nuclear interaction and correlations~\cite{Piarulli:2022ulk}, this makes the spin-dependent matrix elements much more sensitive to the structure inputs.

The density of the spin-orbit operator
$\mathcal O_{\text{LS}, p}$, multipied by a factor of 10, is shown by the yellow points in Fig. \ref{fig:av18}. For both the NV2+3-Ia and AV18+UX models, the LS matrix element turns out to be negligible. This is due to a cancellation between the contributions of the $\LL$ and $\LL^{\text{CM}}$ terms in Eq. \eqref{eq:Oso}. The same feature is observed in $^{14}$O, after the 
$\LL^{\text{CM}}$ contribution neglected in Ref. \cite{Cirigliano:2024msg} is restored \cite{GandolfiPrivate}, and, to a lesser degree, in medium mass nuclei \cite{NovarioPrivate}.

Finally, we remark that the magnetic matrix elements $M_{\text{GT}, p}$
and $M_{\text{GT}, n}$ depend on the ultraviolet cut-off used in the interactions \cite{Cirigliano:2024msg}, which is different 
in the AV18 and NV2-Ia potentials. The difference could be absorbed by fitting the low-energy constants $g_{V1}^{NN}$ and $g_{V2}^{NN}$ to lattice QCD or model calculations of two-nucleon scattering amplitudes, as has been proposed in the case of neutrinoless double-$\beta$ decay \cite{Davoudi:2020gxs,Cirigliano:2021qko}.   
We calculated the two-body corrections to the  ``weak scattering amplitudes''
 $pp(^1S_0) \rightarrow np(^1S_0) e^+ \nu$ and $pn(^1S_0) \rightarrow nn(^1S_0) e^+ \nu$ 
induced by $\mathcal V^{\rm mag}_{\text{GT},p}$ and $\mathcal V^{\rm mag}_{\text{GT},n}$
and found that, over a large range of momenta, they agree at better than 2\%. Any difference between the AV18 and NV2-Ia calculations of the magnetic matrix elements is thus unlikely to be explained by the contributions of the contact interactions.

In Fig.~\ref{fig:chi} we first compare the operator densities computed with the NV2+3-Ia and NV2+3-Ia* interactions. Because these interactions have the same two-body force and differ only in how one constrains the three-nucleon force, it can provide insight into how the latter influences our calculations. Again, panel (a) contains the transition densities for the spin-independent operators, while the dominant contributions to the energy-independent term of $\bar\delta_{\rm NS}$ are shown in panel (b). Now, we represent the NV2+3-Ia* results with open symbols, and the filled symbols again represent the results obtained when using the NV2+3-Ia interaction. In this case, we again see a shift in the radial dependence of the transition densities when going from one model to another. In particular, some short-distance strength in the NV2+3-Ia* model is shifted to larger interparticle spacings, which is likely due to this interaction having a more repulsive three-nucleon force than its counterpart. Because some densities come with a node, shifting where the strength is located means that the specific three-nucleon fore can also have an effect on the overall matrix element. 

\subsection{Analysis of GFMC results}\label{sec:gfmc}

\begin{table}
\begin{tabular}{|| c | c | c | c ||}
\hline
Model & $r_p$ [fm] & $R_{\rm ch}$ [fm] & $E$ [MeV] \\ \hline
NV2+3-Ia & $2.43(2.32)$ &$2.57(2.46)$ &$-61.7(-66.3)$  \\ \hline
NV2+3-Ia* & $2.56(2.48)$ &$2.69(2.61)$ & $-58.2(-61.0)$ \\ \hline \hline
AV18+UX &$2.63(2.50)$ &$2.72(2.63)$ &$-55.7(-57.6)$   \\ \hline
AV18+IL7 &$2.50(2.42)$ &$2.63(2.56)$ &$-59.7(-63.2)$  \\ \hline \hline
Exp & -- &--  &$-60.3$($-63.0)$  \\ \hline
\end{tabular}
\caption{Static properties of the $^{10}{\rm C}(0^+)~(^{10}{\rm B^*}(0^+;1))$ computed with GFMC for all models under study. Experimental data are from~\cite{Tilley:2004zz}. The GFMC calculations have an $\approx 1\%$ error arising due to statistical uncertainties.}\label{tab:gfmc.static}
\end{table}

In order to provide the most accurate description of $\bar\delta_{\rm NS}$ in the present framework, we performed GFMC propagations for all of the matrix elements. These results are included along with the VMC numbers in Tables~\ref{TabI} and~\ref{TabII}. Additionally, Table~\ref{tab:gfmc.static} summarizes the static properties of the GFMC wave functions. Namely, we present the energies $E$, point proton radii $r_p$, and charge radii $R_{\rm ch}$ for the nuclei under study. We find that our results with the AV18+IL7 provide satisfactory agreement with previous calculations~\cite{Carlson:2014vla}, but the AV18+UX model does not provide sufficient binding for the two systems. For the chiral potentials, we find that the central value for the NV2+3-Ia in this work provides too much binding for the systems when compared to experiment, while the NV2+3-Ia* underbinds. 

To compute $R_{\rm ch}$, we use the formula~\cite{Friar:1997js}
\begin{equation}
    \avg{R_{\rm ch}^2} = r_p^2 + \avg{R_p^2} + \frac{N}{Z} \avg{R_n^2} + \frac{3}{4m_p^2}\, ,
\end{equation}
where $\avg{R_p^2} = 0.7071$ fm$^2$ is the square of the proton charge radius ~\cite{ParticleDataGroup:2024cfk}, $\avg{R_n^2} = -0.1155(17)$ fm$^2$ is the square of the neutron charge radius~\cite{ParticleDataGroup:2024cfk}, and $3/(4m_p^2) \approx 0.033$ fm$^2$ is the Darwin-Foldy correction~\cite{Friar:1997js}. This formula is generally consistent with leading order charge radius extracted from the electric form factor, and because two-body currents provide a percent level correction to this observable~\cite{King:2025akz}, the reported values are qualitatively sufficient to look for correlations with the computed matrix elements.
As expected, the energies and radii are anticorrelated, i.e., the greater the binding, the smaller the radius.
Although there are presently no data with which to compare the computed ground state radius of $^{10}$C, such data could be useful both for benchmarking the GFMC calculations, and for proposed strategies to empirically determine isospin symmetry breaking effects relevant to superallowed $\beta$ decays~\cite{Seng:2022epj}.

\begin{figure}
    \centering
    \includegraphics[width=0.95\linewidth]{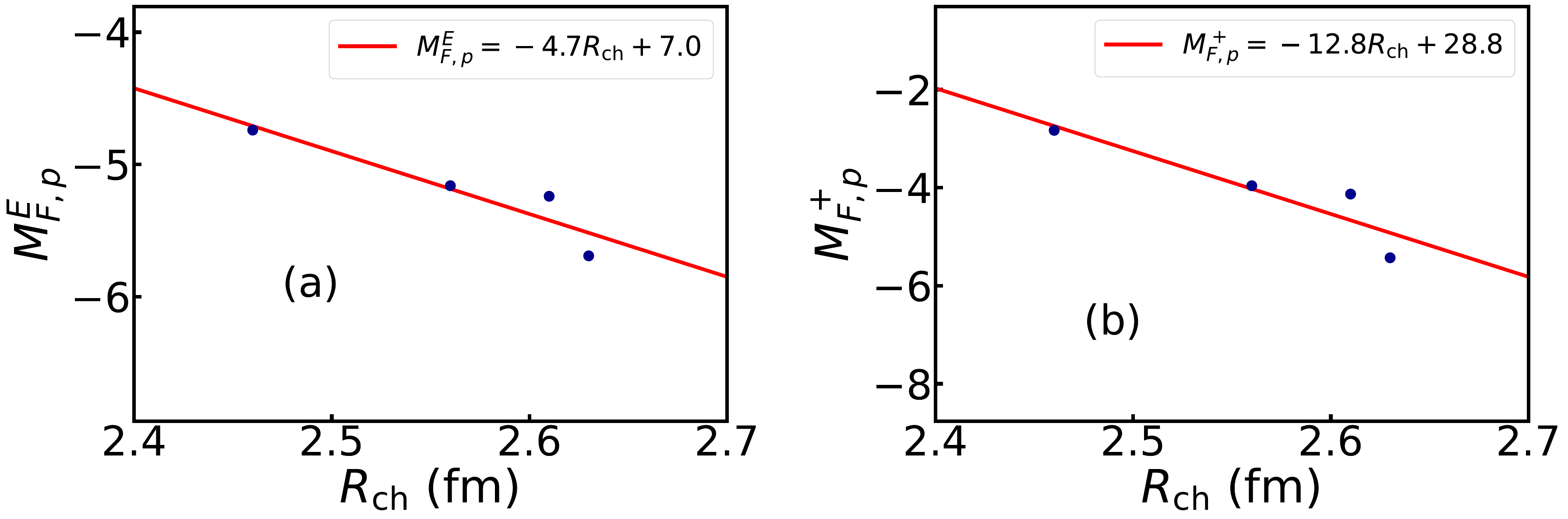}
    \caption{(a) $M^E_{\text{F},p}$ and (b) $M^+_{\text{F},p}$  as a function of the charge radius $R_{\rm ch}$ of $^{10}{\rm B^*}(0^+;1)$ for the GFMC calculations shown by the blue points. The red lines represent linear fits to the calculated values.}
    \label{fig:mefp.vs.rch}
\end{figure}

Despite the spread in energies, we find that none of the matrix elements are strongly impacted by the binding energies of the system. For the dominant spin-dependent matrix elements, we find that there is no pattern of correlation between the interactions and the matrix elements. For the spin-independent matrix elements,  $M^E_{\text{F},p}$ and $M^+_{\text{F},p}$, we instead find that $R_{\rm ch}$ and $R_p$ correlate with these matrix elements. Fig.~\ref{fig:mefp.vs.rch} display $M^E_{\text{F},p}$ and $M^+_{\text{F},p}$ as a function of $R_{\rm ch}$, respectively. Included are lines of best fit given by,
\begin{eqnarray}
    M^E_{\text{F},p} &=& -4.7 R_{\rm ch}(^{10}{\rm B^*}(0^+;1)) + 7.0\, ,\\
    M^+_{\text{F},p} &=& -12.8 R_{\rm ch}(^{10}{\rm B^*}(0^+;1)) + 28.8\, .
\end{eqnarray}

In the standard approach to electromagnetic corrections to superallowed decays, $M_{\text{F},p}^{E}$ is accounted for by terms in the finite-nuclear-size corrections $L_0(Z,E_e)$
and in the shape correction $C(Z,E_e)$,
which depend on the nuclear charge distribution via 
the parameter $R = \sqrt{5/3} \sqrt{\langle R^2_{\rm ch} \rangle}$. $M^+_{\text{F},p}$, on the other hand, would reproduce the dependence of the Fermi function on the nuclear radius \cite{Cirigliano:2024msg}. 
As the charge radius of $^{10}{\rm B^*}(0^+;1)$ is not measured, the radius of the stable $^{10}{\rm B}(3^+)$,
$R_{\rm ch} = 2.428(5)$ fm \cite{Angeli:2013epw}, is typically used \cite{Hardy:2004id}.
In this approach, one would have
\begin{align}
    \tilde{M}^E_{\text{F}, p} &= -\mathcal M_F \frac{ R}{2 R_A} = -4.29, \\
    \tilde{M}^+_{\text{F}, p} &= - \mathcal M_F \log \frac{R}{R_A} = -1.36
\end{align}
$\tilde M^E_{\text{F},p}$ is between 10\% and 25\% smaller than in the \textit{ab initio} calculation. 
Since, as we will see in the next section, $\overline{\delta_{\rm NS}^{E}}$ is about $1 \cdot 10^{-3}$, 
this difference affects the $^{10}$C half-life at the 
$10^{-4}$ level, and it is still negligible. $\tilde M^+_{\text{F},p}$ is even more different from what we obtained, being smaller by up to a factor of 4.
Also in this case, the shift to the half-life is of the order of few $\times 10^{-4}$, and still subdominant for $^{10}$C. We notice, however, that these corrections grow with $Z$, and thus it would be important to control them in medium-mass nuclei.

The GFMC propagations alter the value of $M^{\rm mag}_{\text{GT}, p}$, the dominant contribution to $\bar\delta_{\rm NS}$, by $\lesssim 1\%$ to $\approx 13\%$ across the four calculations. Generally, the values of the matrix elements change by a few percent to order ten percent, with notable exceptions being the spin-independent matrix elements that strongly correlate with the radius, and the matrix element $M_{\text{GT}, n}^{\rm mag}$. In the latter case, we note that similar behavior was observed for QMC calculations of magnetic densities~\cite{Chambers-Wall:2024uhq}. When the matrix element density exhibited a node, it was possible to generate large changes to the integrated value because of sensitivity to the precise details of the spin distribution and the interparticle correlations. Indeed, this operator is similarly spin-dependent, and its matrix element density exhibits a node. Thus, we conclude that it is similarly sensitive to the details of the spin and particle distributions of the two nuclei. 

Finally, we also note that the GFMC propagation does not necessarily enhance or quench the matrix elements uniformly. This is similar to how the inclusion of correlations can reduce or increase the matrix elements of spin-dependent operators, such as in calculations of magnetic moments~\cite{Chambers-Wall:2024uhq,Miyagi:2023zvv}. Once again, such behavior was observed in GFMC calculations when operator densities exhibited nodes, making them dependent on the specifics of the spins and coordinates of nucleons in the system. We therefore conclude that the behavior observed in this study arises for similar reasons. 

\subsection{Effect of many-body correlations}\label{sec:correlations}

\begin{table}
\begin{tabular}{||c| c | c | c |  c | c | c || c | c||}
\hline
Method &  $M^{\rm mag}_{\text{GT}, p}$ & $M^{\rm mag}_{\text{GT}, n}$ & $M^{\rm mag}_{\text{T}}$ & $M_{\text{ LS}, p}$& $M^{\rm CT}_{\text{GT}, p}$ & $M^{\rm CT}_{\text{GT}, n}$  & $M_{\text{F}, p}^+$  & $M^E_{\text{F}, p}$ \\ 
\hline 
VMC (w/ OPE corr.) & $-0.46$ & $0.064$ & $0.007$ & $-0.006$ & $0.050$ & 
$0.018$ & $-3.32$ & $-4.91$
\\
\hline 
VMC (w/o OPE corr.) & $-0.36$ & $0.039$ &   $0.003$ & $-0.008$ & $0.037$
& $0.018$ &  $-3.37$ & $-4.94$ \\
\hline
\end{tabular}
\caption{Matrix elements 
entering
$\bar\delta_{\rm NS}$ 
with and without OPE correlations.
}\label{TabIII}
\end{table}

In order to assess the impact that many-body correlations have on the value of $\bar\delta_{\rm NS}$, we performed a VMC calculation using the NV2+3-Ia without the dominant correlations in the wave function given by Eq.~\eqref{eq:psi.t}, similar to the study of neutrinoless double-$\beta$ decay matrix elements in Ref.~\cite{Pastore:2017ofx}. We compare the  matrix elements in Table~\ref{TabIII}. 
First, we remark that the matrix elements of the spin-independent operators -- $M^E_{\text{F},p}$ and $M^+_{\text{F},p}$ -- are only mildly impacted by removing correlations. This is likely due to the fact that these operators should be insensitive to changes in the spin-distribution induced by turning on or off OPE correlations. Further, these are long-range operators and should also be insensitive to changes in the short-distance physics. 

Removing the OPE correlations has a more pronounced effect on the spin-dependent matrix elements. This is unsurprising, since these operators should depend on the details of the spin distribution, which will be impacted by removing tensor correlations in the wave function. Interestingly, removing OPE correlations quenches the magnitude of the matrix elements. As in the discussion of quenching when going from VMC to GFMC, we attribute this to the fact that the transition densities display nodes. 

We can sum the long-range GT, T and LS~matrix elements
to obtain a quantity that can be more easily compared with the results of Ref.~\cite{Hardy:2020qwl} 
\begin{equation}
    \left. \delta^{(0)}_{\rm NS} \right|_{\rm w/o~OPE~corr.} = - (3.23 + 0.08 ) \cdot 10^{-3} \, ,
\end{equation}
where the first term 
is given by contributions proportional to the nucleon magnetic moment and the second by the spin-orbit term.
These two terms in the chiral transition operators correspond to $\delta_{\rm NS, B}$ in the approach of Ref.~\cite{Towner:1992xm}, with the difference that the short-range behavior of the matrix elements is in that case regulated by the axial and magnetic form factors. The authors of Ref.~\cite{Hardy:2020qwl} find
\begin{equation}
\label{eq:deltaNSB}
    \left. \delta_{\rm NS, B}\right| = - 3.06(35) \cdot 10^{-3},
\end{equation}
which agrees very well with the results we obtain neglecting OPE correlations.

\section{Results for $\bar\delta_{\rm NS}$ and comparison with previous literature }\label{sec:delta.ns}

\begin{table}
\begin{tabular}{||c| c | c | c | c |  c | c | c || c||}
\hline
Model & Method & $\delta_{\rm NS}^{(0)}$  &  $\overline{\delta_{\rm NS}^E}$ \\ 
\hline 
NV2+3-Ia  & VMC & $ -\left( 4.03 + 0.25 \pm 0.69\right) \cdot 10^{-3} $  & $1.01 \cdot 10^{-3}$
\\
&GFMC & $-\left( 4.43+ 0.21 \pm 0.77\right) \cdot 10^{-3}$ &  $0.97 \cdot 10^{-3}$  \\
\hline 
NV2+3-Ia$*$ & VMC & $-\left( 4.18 + 0.31 \pm 0.62 \right) \cdot 10^{-3}$ & $1.06 \cdot 10^{-3}$ \\
&GFMC &  $-\left(4.25 + 0.31  \pm 0.66\right) \cdot 10^{-3}$  & $1.08 \cdot 10^{-3}$ \\
\hline
\hline 
AV18+UX & VMC & $-\left(4.48 + 0.27 \pm 0.67\right) \cdot 10^{-3}$ & $1.02 \cdot 10^{-3}$ \\
&GFMC & $-\left( 4.06 +0.40 \pm 0.48\right) \cdot 10^{-3}$ & $1.17 \cdot 10^{-3}$\\ 
\hline
AV18+IL7 & VMC & $-\left( 4.55 + 0.26 \pm 0.61 \right) \cdot 10^{-3}$ & $1.01 \cdot 10^{-3}$ \\
&GFMC &$-\left( 4.32 + 0.29 \pm 0.64\right) \cdot 10^{-3}$ & $1.06 \cdot 10^{-3}$\\ \hline
\end{tabular}
\caption{Summary of the results for $\bar\delta_{\rm NS}$.
For the energy-independent component, $\delta_{\rm NS}^{(0)}$,
the first term encodes the contribution of the magnetic and spin-orbit two-body operators. The second term is the $\mathcal O(\alpha^2)$ correction arising from $M_{\text{F}, p}^+$. The last contribution arises from the contact interactions, with LECs set to the {\it  arbitrary} values $g_{V1}^{NN} \pm g^{NN}_{V2} = \pm \frac{1}{m_N} \frac{1}{(2 F_\pi)^2}$. The last column shows the energy dependent part, $\overline{\delta_{\rm NS}^{\rm E}}$. 
}\label{Tab:deltaNS0}
\end{table}

We can combine the results shown in Table \ref{TabI} 
to obtain values for $\delta^{(0)}_{\rm NS}$
and $\overline{\delta^{\text{E}}_{\rm NS}}$,
which we give in Table \ref{Tab:deltaNS0}.
To facilitate the comparison with previous work, we separate the contributions to $\delta_{\rm NS}^{(0)}$ that arise from the long-range magnetic and spin-orbit operators (first term in the third column in Table \ref{Tab:deltaNS0}), from the $\mathcal O(\alpha^2)$ correction
induced by $M_{\text{F},p}^+$ (second term)
and the short-range contributions proportional to the LECs $g^{NN}_{V1}$
and $g_{V2}^{NN}$ (third term). 
Since the LECs are not known, for this last contribution we use the dimensional analysis estimate
\begin{equation}
    g_{V1}^{NN} \pm g^{NN}_{V2} = \pm \frac{1}{m_N} \frac{1}{(2 F_\pi)^2},
\end{equation}
and report the ensuing range of values.

We can compare the 
long-range part of $\delta_{\rm NS}^{(0)}$ 
with the shell-model calculation used
in Ref.~\cite{Hardy:2020qwl}.
Focusing on the GFMC results, we find that, for the four interactions we consider, the value of $\delta^{(0)}_{\rm NS}$ falls within the range
\begin{equation}
   \left. \delta_{\rm NS}^{(0)} \right|_{\rm mag + LS} = - \left[4.06, 4.43\right] \cdot 10^{-3}.
\end{equation}
The magnetic and spin-orbit operators have a close correspondence to contribution to $\delta_{\rm NS, B}$.
The value we find is about 40\% larger than Eq. \eqref{eq:deltaNSB}, and deviates from it by about $3\sigma$. 
As discussed above, most of the deviation can be attributed to the additional correlations included in the GFMC wave function. 

The value of $\delta_{\rm NS}$ adopted in Ref.~\cite{Hardy:2020qwl},\begin{equation}\label{eq:deltaNStot}
    \delta_{\rm NS} =
    \delta_{\rm NS, A} + \delta_{\rm NS, B} +\delta_{\rm NS, E}   =
    -4.0(5) \cdot 10^{-3},
    \end{equation}
includes additional corrections arising from the modification of the nucleon $W \gamma$ box in nuclei and from the energy-dependent correction identified in Ref.~\cite{Gorchtein:2018fxl}.  
At least part of these corrections are due to short-range physics and are captured by the short-distance piece of $\delta_{\rm NS}^{(0)}$. We see that, within the fairly large range of LECs we adopt, our value is compatible with Eq. \eqref{eq:deltaNStot}.

Ref.~\cite{Gennari:2024sbn}
presented a calculation of $\delta_{\rm NS}$ based on the 
evaluation of the axial-electromagnetic Compton amplitude $T_3(\nu, {\bf q})$ 
using the no-core shell model (NCSM)~\cite{Barrett:2013nh} to solve the many-body problem, together with chiral Hamiltonians, weak and electromagnetic currents. The calculation has many similarities with the approach of Ref.~\cite{Cirigliano:2024msg}, with the main difference that contributions to $T_3(\nu, {\bf q})$ are not expanded using the method of regions, and 
possible sensitivity to the region $|{\bf q }| \gg k_F$
which arises when performing the integral over $\bf q$
is not absorbed into  LECs.
The authors of Ref.~\cite{Gennari:2024sbn} find 
\begin{equation}\label{eq:deltaNSnocore}
    \delta_{\rm NS} = -\left(4.22 \pm 0.32 \right) \cdot 10^{-3}.
\end{equation}
If we consider the contributions of the LECs as theoretical error, our results are very close in central value and have a larger theoretical error compared to  Eq.~\eqref{eq:deltaNSnocore}. The two results are compatible within error. It would be interesting to understand how much of the difference arises from long-range physics, $|{\bf q}| \lesssim  k_F$, and could thus be attributed to the many-body methods and nuclear interactions, and how much 
to short-range physics, and thus could be
absorbed by the values of the LECs.
This question could be addressed by fitting the LECs to data, or by estimating them by modeling two-nucleon amplitudes \cite{Cirigliano:2020dmx,Cirigliano:2021qko}.

Secondly we consider the $\mathcal O(\alpha^2)$ correction to $\delta_{\rm NS}$ induced by $M_{\text{F}, p}^+$. The GFMC results yield
\begin{equation}
    \left. \delta^{(0)}_{\rm NS} \right|_{\alpha^2} = - [0.21,0.40] \cdot 10^{-3}.
\end{equation}
This contribution is usually included in the phase-space factor, via the logarithmic dependence of the Fermi function on $R$. We can estimate the impact on the half-life by considering the difference we would obtain using the ``naive'' matrix element $\tilde M^+_{F, p}$  \begin{equation}
    \left. \Delta \delta^{(0)}_{\rm NS} \right|_{\alpha^2} 
    = - [0.11,0.30] \cdot 10^{-3}.
\end{equation}
We thus obtain a deviation of few $\times 10^{-4}$, which, as we will see, is still subleading with respect to other theoretical errors.

We show the results for the matrix elements that enter $\overline{\delta_{\rm NS}^E}$ in the last column of Table \ref{Tab:deltaNS0}. As already noticed in Ref.~\cite{Cirigliano:2024msg}, the dominant correction arises from $M^E_{F, p}$, while pion matrix elements are small, and we neglect them. Under this assumption, we obtain,
\begin{equation}
  \overline{  \delta_{\rm NS}^E} = [0.97, 1.17] \cdot 10^{-3} \, .
\end{equation}
This correction is usually accounted for by terms in the finite-nuclear-size corrections $L_0(Z,E_e)$
and in the shape correction $C(Z,E_e)$.
We can get an idea of the shift in the half-life by computing the difference between $\overline{\delta_{\rm NS}^E}$ 
and what one would obtain using the naive matrix element $\tilde M_{\text{F},p}^E$
\begin{equation}
    \Delta \delta_{\rm NS}^E = [0.09,0.19] \cdot 10^{-3},
\end{equation}
which, once again, is subdominant.

Finally, we assess the impact of using different nuclear models on the extraction of $V_{ud}$. 
We can extract $V_{ud}$ using the master formula 
\begin{align}
    \frac{1}{t}  &=    \frac{G_F^2 |V_{ud}|^2  m_e^5}{\pi^3 \log 2}\,  
    \Big[C_\text{eff}^{(g_V)}(\mu)\Big]^2  \quad\times[1 + \bar \delta_R^\prime(\mu)]\,  (1 +  \bar \delta_\text{NS}) \,  ( 1 - \bar  \delta_C)  \,  \bar f(\mu).
    \label{eq:master2}
\end{align}
Here $t$ denotes the partial half-life given in Ref.~\cite{Hardy:2020qwl}
\begin{equation}
    t = 1321.8 \pm 1.8 \, \text{s},
\end{equation}
while the prefactor on the right-hand-side 
\begin{equation}
    \frac{G_F^2 m_e^5}{\pi^3 \log 2} = 3.350722(3) \times 10^{-4}\,\text{s}^{-1},
\end{equation}
is known with negligible uncertainty.
Using the recipe in Ref.~\cite{Cirigliano:2024msg}, 
we find the combination $ \left[C_\text{eff}^{(g_V)}\right]^2\bar f (1 + \bar\delta_R^\prime)$ to be
\begin{equation}\label{combscale0}
   \left[C_\text{eff}^{(g_V)}\right]^2\bar f (1 + \bar\delta_R^\prime) = 2.39519 (56)_{g_V} (87)_\mu,
\end{equation}
where the first error is due to the nonperturbative uncertainty in the nucleon $W\gamma$ box ~\cite{Cirigliano:2023fnz,Seng:2018qru,Seng:2018yzq,Czarnecki:2019mwq,Shiells:2020fqp,Hayen:2020cxh,Seng:2020wjq,Cirigliano:2022yyo}.
The second error is obtained by varying the low-energy renormalization scale $\mu$ between $E_0$ and $4 E_0$. This variation estimates some missing $\mathcal O(\alpha^2 Z)$ terms in the phase space factor $\bar f$.
For the isospin breaking correction $\delta_C$,  we assume the result from Ref.~\cite{Hardy:2020qwl} 
\begin{equation}
   \overline{\delta}_C = \left(1.75 \pm 0.18\right) \cdot 10^{-3},
\end{equation}
even though this quantity should in principle be computed using the same many-body method and chiral interactions as $\bar\delta_{\rm NS}$. 

Putting these factors together, we find that, using only the $^{10}$C measurement, one would get the following values of $V_{ud}$
\begin{align}
 & \text{NV2+3-Ia}     &\left. V_{ud} \right|_{^{10} \rm C} &= 0.97355 (66)_{\rm exp} (12)_{g_V}  (17)_{\mu} (9)_{\delta_C}  (38)_{g_V^{\rm NN}}, \\
 & \text{NV2+3-Ia*}     &\left. V_{ud} \right|_{^{10} \rm C} &= 0.97345 (66)_{\rm exp} (12)_{g_V}  (17)_{\mu} (9)_{\delta_C}  (32)_{g_V^{\rm NN}},
\\
&\text{AV18+UX}  & \left. V_{ud} \right|_{^{10} \rm C} &= 0.97336 (66)_{\rm exp} (12)_{g_V}  (17)_{\mu} (9)_{\delta_C}  (23)_{g_V^{\rm NN}} , \\
&\text{AV18+IL7} & \left. V_{ud} \right|_{^{10} \rm C} &= 0.97349 (66)_{\rm exp} (12)_{g_V}  (17)_{\mu} (9)_{\delta_C}  (31)_{g_V^{\rm NN}}.
\end{align} 
The four different interactions we use result in a spread of $1.9 \cdot 10^{-4}$, comparable with the uncertainty due to $g_V$ and missing $\mathcal O(\alpha^2 Z)$ terms, but smaller than the effect due to the missing low-energy constants (denoted by $g_V^{\rm NN}$). 
For comparison, using $^{10}$C only and 
the theoretical corrections discussed in Ref.~\cite{Hardy:2020qwl}, one would get
\begin{equation}\label{eq:VudTH}
    \left. V_{ud} \right|^{\rm HT}_{^{10} \rm C} = 0.97318(66)_{\rm exp}(9)_{\Delta_R^V}(24)_{\delta_{\rm NS}}(9)_{\delta_C},
\end{equation}
while using $\delta_{\rm NS}$ and $\Delta^V_{R}$
from Ref. \cite{Gennari:2024sbn} 
would yield
\begin{equation}\label{eq:VudGennari}
    \left. V_{ud} \right|^{\small \cite{Gennari:2024sbn}}_{^{10} \rm C} = 0.97317(66)_{\rm exp}(9)_{\Delta_R^V}(16)_{\delta_{\rm NS}}(9)_{\delta_C}.
\end{equation}
Even neglecting the large experimental error, 
the range of values we find is compatible with 
Eqs. \eqref{eq:VudTH} and \eqref{eq:VudGennari}.

\section{Conclusion}\label{sec:conclusions}

This work details the calculation of the nuclear structure correction $\bar\delta_{\rm NS}$ to $^{10}{\rm C}$ superallowed $\beta$ decay using state-of-the-art nuclear many-body approaches with an EFT framework. The formalism, developed in Refs.~\cite{Cirigliano:2023fnz,Cirigliano:2024msg,Cirigliano:2024rfk}, allowed for the description of $\bar\delta_{\rm NS}$ in terms of nuclear matrix elements of two-body currents. We employed two methods -- VMC and GFMC -- to study both the values of the matrix elements, as well as their sensitivities to the underlying nuclear dynamics. We presented a range of values for $\bar\delta_{\rm NS}$ arising from the four models of nuclear interaction used in this work. 

Our results can be compared with both the standard value of the review in Ref.~\cite{Hardy:2020qwl}, as well as the recent NCSM calculation of Ref.~\cite{Gennari:2024sbn} based on the dispersion formalism of Refs.~\cite{Seng:2018qru,Seng:2018yzq,Gorchtein:2018fxl,Seng:2020wjq,Seng:2022cnq,Gorchtein:2023naa}. Treating the unknown low-energy constants as a theoretical uncertainty on the GFMC results, we find that our number agrees with the other evaluations within error. Our central values are in very good agreement with that of Ref.~\cite{Gennari:2024sbn}, but higher than Ref.~\cite{Hardy:2020qwl}. The current theoretical uncertainty on our approach is larger than that obtained by the other recent {\it ab-initio} evaluation in Ref.~\cite{Gennari:2024sbn}; however, this could be improved with a determination of the unknown low-energy constants, requiring a calculation of $\bar\delta_{\text{NS}}$ in multiple isotopes with the same nuclear Hamiltonian. Our error estimate reflects the uncertainty of using a small class of chiral interactions, and may under- or over-estimate the true uncertainty from a full exploration of the parameter space. With both the developments of order-by-order local chiral interactions~\cite{Bub:2024gyz,Somasundaram:2023sup} and emulators for QMC calculations~\cite{Somasundaram:2024zse,Armstrong:2025tza}, one could make a full quantification of the nuclear uncertainties on $V_{ud}$. This work thus represents a first step toward a robust determination of $V_{ud}$ that will allow us to meaningfully evaluate the current status of CKM unitarity. 

In addition to computing $\bar\delta_{\rm NS}$, we also extracted $V_{ud}$ from the available experimental data for this transition. Our spread of models provides an uncertainty on the order of $2\cdot 10^{-4}$, which is comparable with the uncertainty arising due to $g_V$, but smaller that the uncertainty due to the unknown low-energy constants. When accounting for these effects, the dominant source of uncertainty is from experiment. If we compare to the extraction of $V_{ud}$ using only the $^{10}{\rm C}$ decay and the theoretical corrections of Refs.~\cite{Hardy:2020qwl,Gennari:2024sbn}, the extractions agree within error and have rather consistent central values. The uncertainty on the GFMC number is larger due, again, to the uncertainty in the low-energy constants. 

While this calculation focused only on the correction $\bar\delta_{\rm NS}$, the isospin-breaking correction $\delta_C$ is also crucial for a determination of $V_{ud}$ using {\it ab initio} nuclear approaches. Thus, this computation will be an important future step in providing an extraction of this parameter that is both high-quality and uses consistent nuclear structure information.
It will also be important to investigate how rigorous is the separation between $\delta_C$ and $\bar\delta_{\rm NS}$. For example, the pion-range contributions to $\bar\delta_{\rm NS}$ calculated in Appendix~\ref{app:a3} depend on the definition of isospin limit used in the nucleon-nucleon potential and two-body currents. The interplay between these two quantities, especially at higher orders in the chiral expansion, remains to be thoroughly investigated.   

$^{10}{\rm C}$ also represents only one of several nuclei that undergo superallowed $\beta$ decay and connect with $V_{ud}$. Calculations of this quantity with methods that are applicable to heavier systems will be important, and this work can serve as a benchmark for such calculations. Finally, the low-energy constants in the contact contributions are the largest source of theoretical uncertainty on the nuclear physics input. Having a way to determine these parameters, either from a fit to superallowed lifetimes, or via lattice QCD approaches or from models based on two-nucleon amplitudes, will help to improve the precision that one can achieve in the EFT approach.

\begin{acknowledgments}
We thank Chien-Yeah Seng for valuable discussions and for pointing out a 
sign mistake in the Fourier transform of the spin-orbit term of the two-body transition operator given in Ref. \cite{Cirigliano:2024msg}. 
We thank S. Novario for several useful discussions, for sharing preliminary results on coupled-cluster calculations of $\delta_{\rm NS}$ and for pointing out missing center-of-mass terms in the spin-orbit operator.
We thank V.~Cirigliano, W. Dekens, J.~de~Vries and  M.~Hoferichter for discussions and comments on the manuscript.
This work is supported by the US Department of Energy under Contracts No. DE-SC0021027 (S.P.) and DE-AC02-06CH11357 (R.B.W.) and the FRIB Theory Alliance under Award No. DE-SC0013617 (A.R.F. and M.P.); and the Office of Advanced Scientific Computing Research, Scientific Discovery through Advanced Computing (SciDAC) NUCLEI program program (J.C., A.R.F., S.G. S.P., M.P., and R.B.W.).
Financial support by Los Alamos
National Laboratory's Laboratory Directed Research and Development program under projects
20220672DR (S.~G. and E.~M.)
and 20240742PRD1 (G.~B.~K.)
is gratefully acknowledged.  Los Alamos National Laboratory is operated by Triad National Security, LLC,
for the National Nuclear Security Administration of U.S.\ Department of Energy (Contract No.\
89233218CNA000001). 
We acknowledge support from the DOE Topical Collaboration ``Nuclear Theory for New Physics,'' award No.\ DE-SC0023663. The work of S.G.\ is also supported by the Network for Neutrinos, Nuclear Astrophysics, and Symmetries (N3AS).
This research used resources provided by the Los Alamos National Laboratory Institutional Computing Program and the Laboratory Computer Resource Center of Argonne National Laboratory.
\end{acknowledgments}

\appendix

\section{Two-body operators in coordinate space}\label{app:potentials}

In this Appendix, we give the expression for the
two-body operators $\mathcal V^{\mathcal X}$ that enter the calculation of $\bar\delta_{\rm NS}$. The operators were originally derived in Ref.~\cite{Cirigliano:2024msg}. We here clarify some of the notation that was left implicit in Ref.~\cite{Cirigliano:2024msg}. We further correct a sign mistake in the coordinate-space version of the spin-orbit potential, and include a term proportional to the two-nucleon center-of-mass momentum that was neglected in Ref.~\cite{Cirigliano:2024msg}. 
We write the two-body electroweak operator  as
the sum of Fermi (F), Gamow--Teller (GT),  tensor (T)  and spin-orbit (LS) components 
\begin{equation}
\label{V_O}
    \mathcal V^\mathcal X = \left(\frac{e^2}{4\pi} \right)^m \sum_{N = p,n} \left( \mathcal V^{\mathcal O} _{\text{F}, N}(\rr) + \mathcal V^{\mathcal O}_{\text{GT}, N}(\rr) + \mathcal V^{\mathcal O}_{\text{T}, N}(\rr)
    + \mathcal V^{\mathcal O}_{\text{LS}, N}(\rr)
    \right), 
\end{equation}
where we separated the contributions arising from couplings to neutrons and protons. 
Here $m=1$ except for $\mathcal V^+_{F, p}$, for which $m=2$.

For electron emissions, the F, GT,  T and LS components are defined as
\begin{align}
\label{V_O_F_GT_T}
    \mathcal V^{\mathcal O}_{\text{F}, N} &=  \sum_{j < k} h_{\text{F}, N}^{\mathcal O}(r_{jk}) \Big[ \tau^{+ (j)}  P^{(k)}_N  + (j \leftrightarrow k)  \Big], \\
    \mathcal V^{\mathcal O}_{\text{GT}, N} &=  \sum_{j < k} h_{\text{GT}, N}^{\mathcal O}(r_{jk}) \, \bsigma^{(j)} \cdot  \bsigma^{(k)} \Big[ \tau^{+ (j)}  P^{(k)}_N  + (j \leftrightarrow k)  \Big], \notag\\
    \mathcal V^{\mathcal O}_{\text{T}, N} &=  \sum_{j < k} h_{\text{T}, N}^{\mathcal O}(r_{jk}) S^{(jk)}({\bf \hat r}) \Big[ \tau^{+ (j)}  P^{(k)}_N  + (j \leftrightarrow k)  \Big],\notag \\
        \mathcal V^{\mathcal O}_{\text{LS}, N} &=  \sum_{j < k} h_{\text{LS}, N}(r_{j k}) \Big[ \tau^{+ (j)} P_N^{(k)}  (\LL_{jk} - \LL^{\text{CM}}_{jk} ) \cdot \boldsymbol{\sigma}^{(j)}  + (j \leftrightarrow k) \Big],\notag
\end{align}
while for positron emission (which involves the great majority of superallowed decays), we have
\begin{align}
\label{V_O_F_GT_T}
    \mathcal V^{\mathcal O}_{\text{F}, N} &=  \sum_{j < k} h_{\text{F},N}^{\mathcal O}(r_{jk}) \Big[ \tau^{- (j)}  P^{(k)}_N  + (j \leftrightarrow k)  \Big], \\
    \mathcal V^{\mathcal O}_{\text{GT}, N} &=  \sum_{j < k} h_{\text{GT}, N}^{\mathcal O}(r_{jk}) \, \bsigma^{(j)} \cdot  \bsigma^{(k)} \Big[ \tau^{- (j)}  P^{(k)}_N  + (j \leftrightarrow k)  \Big], \notag\\
    \mathcal V^{\mathcal O}_{\text{T}, N} &=  \sum_{j < k} h_{\text{T},N}^{\mathcal O}(r_{jk}) S^{(jk)}({\bf \hat r}) \Big[ \tau^{- (j)}  P^{(k)}_N  + (j \leftrightarrow k)  \Big],\notag
    \\
        \mathcal V^{\mathcal O}_{\text{LS}, N} &=  \sum_{j < k} h_{\text{LS}, N}(r_{j k}) \Big[ \tau^{- (j)} P_N^{(k)}  (\LL_{jk} - \LL^{\text{CM}}_{jk} )  \cdot \boldsymbol{\sigma}^{(j)}  + (j \leftrightarrow k) \Big],\notag 
\end{align}
In these expressions, $r_{jk}= |\rr_j - \rr_k|$ and \begin{align}
S^{(jk)}(\hat \rr)&=  3 \hat\rr \cdot \bsigma^{(j)} \, \hat\rr  \cdot \bsigma^{(k)} - \bsigma^{(j)} \cdot \bsigma^{(k)}, \\
\LL_{jk} &= - \frac{i}{2} \rr_{jk}\times \left({\boldsymbol{\nabla}}_j - {\boldsymbol{\nabla}}_k\right),
\\
\LL^{\text{CM}}_{jk} &= - \frac{i}{2} \rr_{jk}\times \left({\boldsymbol{\nabla}}_j + {\boldsymbol{\nabla}}_k\right)
.
\end{align}
Ref.~\cite{Cirigliano:2024msg} neglected the term proportional to $\LL^{\text{CM}}_{jk}$, which vanishes in the two-nucleon center-of-mass frame.   
We then define the matrix elements
\begin{align}
M^{\mathcal X}_{i, N} &=\int_0^\infty d r   \, C^{\mathcal X}_{i, N}(r) =    \langle f |  \mathcal V^{\mathcal X}_{i, N} | i \rangle, \label{eq:density}
\end{align}
where $i = \{\text{F} ,\text{GT}, \text{T}, \text{LS}\}$. Notice that the factors of $e^2/(4\pi)$ are stripped out of the matrix elements and are accounted for directly in $\delta_{\rm NS}^{(0)}$ and $\delta_{\rm NS}^{\rm E}$.

\subsection{Operators contributing to the energy-independent correction $\delta^{(0)}_{\rm NS}$}

At $\mathcal O(\alpha)$,  $\bar\delta_{\rm NS}$ receives contributions from magnetic, contact and spin-orbit operators. The radial functions are
\begin{align}
    h^\text{mag}_{\text{GT}, p}(r) 
    &= 4 h^\text{mag}_{T, p}(r)
    = \frac{g_A}{3 \mN}   \frac{1+\kappa_p}{r}, \\ 
    h^\text{mag}_{\text{GT}, n}(r) &=4 h^\text{mag}_{\text{T}, n}(r) 
     = \frac{g_A}{3 \mN}   \frac{\kappa_n}{r}, \\
     h^\text{CT}_{\text{GT}, p}(r) &= - \frac{4\pi}{3} (g^{N\!N}_{V1} + g^{N\!N}_{V2}) \delta_{R_S}(r), \notag\\  h^\text{CT}_{\text{GT}, n}(r) &= - \frac{4\pi}{3} (g^{N\!N}_{V1} - g^{N\!N}_{V2}) \delta_{R_S}(r) \\
     h_{\text{LS}, p}(r) &= + \frac{g_A}{2 m_N} \frac{1}{r} \\
     h_{\text{LS}, n}(r) &= 0.
\end{align}
Here $g_A = 1.27$, $\kappa_p$ and $\kappa_n$
are the proton and neutron anomalous magnetic moments, $\kappa_p = 1.79$ and $\kappa_n = -1.91$. 
Notice that the sign of the spin-orbit radial function differs from Ref.~\cite{Cirigliano:2024msg}.
$\delta_{R_S}$ is a regularization of the delta function, which needs to be chosen consistently with the strong potential. For the NV2+3-Ia potentials we use,
\begin{equation}\label{eq:shortrange}
    \delta^{(3)}(\rr) \rightarrow \delta_{R_S}(r) =\frac{1}{\pi^{3/2}  R_S^3} \exp\left(-\frac{r^2}{R^2_S}\right),
\end{equation}
with $R_S=0.8$ fm$^{-1}$, which is consistent with the regulator used in the chiral interaction. For the AV18, we opt to use the same radial function as in Eq. \eqref{eq:shortrange} to smear the delta functions. While, in principle, this choice is inconsistent with the short range dynamics of the AV18 potential, it allows us to compare the effect of changing the model on the overall matrix element more directly. Should it become possible to determine the LECs entering these terms with more precision, this contact could be revisited with more care given to the consistency in the short-range radial function.

The LECs $g_{V1}^{NN}$
and $g_{V2}^{NN}$ have dimension $\left[-3\right]$. For obtaining dimensionless results, we find it convenient to rescale
\begin{equation}
    g_{V j}^{NN} = \frac{1}{m_N} \frac{1}{(2 F_\pi)^2} \tilde{g}_{Vj}^{NN}.
\end{equation}
In our calculation we set $\tilde g_{Vj}^{NN} = \pm 1$.

At $\mathcal O(\alpha^2)$ there is a further two-body contribution
\begin{align}
    h^+_{{\rm F}, p}(r,\Lambda) &= - \log(r \Lambda), \\
    h^+_{{\rm F}, n}(r,\Lambda) &= 0. 
\end{align}
This depends on a cut-off $\Lambda$. 
This dependence is canceled by our treatment of the Fermi function. We evaluate the matrix element for the choice $\Lambda = R_A^{-1}$, with $R_A = 1.2 A^{1/3}$ fm. 
Considering these contributions, one gets
\begin{align}\label{deltaNS0app}
    \delta^{(0)}_\text{NS}  &= \frac{2}{g_V(\mupi)M_\text{F}^{(0)}}   \sum_{N = n,p} \bigg[ \alpha  \big( M^\text{mag}_{\text{GT}, N} + M^\text{mag}_{\text{T}, N} +  M^\text{CT}_{\text{GT}, N}    + M_{\text{LS}, N}  \big)      
    + \alpha^2  M^+_{\text{F}, N}\bigg] .
\end{align}

At $\mathcal O(\alpha^2)$, there is also a three-body component
\begin{align}
   C_+^\text{3b} \,    {\mathcal V}_+^\text{3b}(\rr, \Lambda) &=  - g_V(\mu)\frac{\alpha^2}{2} \times\sum_{i \neq j \neq k}  \log \bigg[\frac{\Lambda}{2} \Big(r_{ij} + r_{ik} + r_{jk} \Big) \bigg] \tau^{\pm(i)}  P_p^{(j)} P_p^{(k)},
    \label{eq:V3plusMain}
\end{align}
which we have not evaluated.

\subsection{Operators contributing to the energy-dependent correction $\overline{\delta^E_{\rm NS}}$}

The energy-dependent correction receives the following contributions
\begin{align}\label{VEpir}
    h^E_{\text{F}, p}(r)  &= - \frac{r}{2 R_A},\notag\\
        h^E_{\text{F}, n}(r)  &= 0,\notag\\
    h^{E \pi}_{\text{GT}, p}(r)  &= - h^{E \pi}_{\text{GT}, n}(r) = \frac{g^2_A Z_\pi}{3 }
     \frac{e^{- \mpi r}}{72 \mpi R_A}  \big(12 + 12 \mpi r - \mpi^2 r^2 \big), \notag\\
     h^{m_e \pi}_{\text{GT}, p}(r)  &= - h^{m_e \pi}_{\text{GT}, n}(r) = \frac{g^2_A Z_\pi}{3 }
     \frac{e^{- \mpi r}}{72 \mpi R_A}  \big(15 - 21 \mpi r + \mpi^2 r^2 \big), \notag\\
     h^{E\pi}_{\text{T}, p}(r) &=-h^{E\pi}_{\text{T}, n}(r) = \frac{g^2_A Z_\pi}{3}\frac{e^{- \mpi r}}{72 \mpi R_A }  \big( 9 \mpi r - \mpi^2 r^2\big),\notag\\
     h^{m_e\pi}_{\text{T}, p}(r) &=-h^{m_e\pi}_{\text{T}, n}(r) = -\frac{g^2_A Z_\pi}{3}\frac{e^{- \mpi r}}{72 \mpi R_A }  \big( 18 \mpi r - \mpi^2 r^2\big).
\end{align}
In order to define dimensionless matrix elements, we normalized all radial functions by $R_A$.
Here $M_\pi = M_{\pi^0}$, and $Z_\pi$ is determined from the pion mass splitting
\begin{equation}
    M_{\pi^\pm}^2 - M_{\pi^0}^2 = 2 e^2 F_\pi^2 Z_\pi, 
\end{equation}
implying $Z_\pi \sim 0.8$.
The pion-range contributions depend on the definition of the isospin limit and how the isospin limited is implemented in standard two-body weak currents and in the nucleon-nucleon potential. In Ref. \cite{Cirigliano:2024msg}, the isospin limit was defined by $M_\pi = M_{\pi^0}$. The functions $h^{E\pi}_{\text{GT}, N}$,
$h^{E\pi}_{\text{T}, N}$,
$h^{m_e \pi}_{\text{GT}, N}$ and $h^{m_e \pi}_{\text{T}, N}$ will  change if a different definition of isospin limit is implemented in two-body corrections to the weak charge operator and to the nucleon potential.

The energy-dependent $\mathcal O(\alpha \epsilon_{\slashed{\pi}})$ corrections are then given by
\begin{align}\label{deltaNSEapp}
\overline{\delta^E_\text{NS}} &= \mp \alpha \frac{2}{g_V(\mupi)M_\text{F}^{(0)}} R_A E_0
\sum_{N=n,p} \bigg[
\tilde{f}_E  M^E_{\text{F}, N}  + 
 \left( M^{E\pi}_{\text{GT}, N}+ M^{E\pi}_{\text{T}, N}\right)+
\tilde f_{m_e}^\pi
 \left( M^{m_e\pi}_{\text{GT}, N}+ M^{m_e\pi}_{\text{T}, N}\right)\bigg],
\end{align}
where the upper sign is for $\beta^+$, the lower sign for $\beta^-$ decays.
$E_0$ is the electron endpoint energy 
\begin{equation}
    E_0 = \mathcal Q_{\rm EC} - m_e,
\end{equation}
in terms of the electron capture 
$\mathcal Q_{\rm EC}$ value given in Ref.~\cite{Hardy:2020qwl}.
The factors 
$\tilde f_E$
and $\tilde f_{m_e}^\pi$
arise from the phase-space average, and are given by
\begin{align}
 \tilde{f}_E &= \frac{1}{E_0}   \left(\frac{4}{3}\langle E_e \rangle + \frac{1}{6} E_0 + \frac{1}{2}\left\langle \frac{m^2_e}{ E_e} \right\rangle \right),\notag\\
 \tilde{f}_{m_e}^\pi &=    \frac{1}{E_0}\left\langle \frac{m^2_e}{E_e} \right \rangle, 
\end{align}
with 
\beq
 \langle E_e^n \rangle = \frac{\int_{m_e}^{E_0} d E_e w_0(E_e)  
    \, \tilde C (E_e) \, 
 \bar F (\beta, \mu) \,  
    E_e^n }{\int_{m_e}^{E_0} d E_e  w_0 (E_e)\, \tilde C (E_e) \, 
 \bar F (\beta, \mu)  }~.
\label{eq:PSaverage2}
 \eeq 
$w_0$ is a kinematical factor
\begin{equation}
    w_0 = |\vec p_e| E_e (E_0 - E_e)^2,
\end{equation}
$\tilde C$ encodes corrections that can be neglected when evaluating $\delta_{\rm NS}^{\rm E}$, $\tilde C = 1$. $\bar F$ is the EFT Fermi function
 \begin{align}\label{eq:FF}
\bar F(\beta,\mu)  &= \frac{4\eta}{(1+\eta)^2}\frac{2(1+\eta)}{\Gamma(2\eta+1)^2}|\Gamma(\eta+iy)|^2 e^{\pi y} \left(\frac{2 |\pp_e|}{\mu}e^{1/2-\gamma_{E}}\right)^{2(\eta-1)},
\end{align}
with $\eta = \sqrt{1-\alpha^2 Z^2}$ and $y =\mp Z\alpha/\beta$. Here $Z$ denotes the charge of the final state nucleus.

\subsection{Pion-range contributions to $\overline{\delta_{\rm NS}^{E}}$}
\label{app:a3}

\begin{table}
\begin{tabular}{||c| c | c | c |  c | c ||}
\hline
Model & Method &   $M^{E \pi}_{\text{ GT}, p}$ & $M^{E \pi}_{\text{GT}, n}$ & $M^{m_e \pi}_{\text{GT}, p}$ & $M^{m_e \pi}_{\text{GT}, n}$ \\ 
\hline 
NV2+3-Ia &VMC &   $-0.048$ & $0.014$ & $0.013$ & $-1.4 \cdot 10^{-3}$ \\
&GFMC & $-0.053$ &$0.018$ &$0.016$ & $-1.5\times 10^{-3}$\\
\hline 
NV2+3-Ia$*$ & VMC   & $-0.048$ &   $0.013$ & $0.016$ & $-1.7 \cdot 10^{-3}$ \\
&GFMC &   $-0.049$ & $0.017$ & $0.017$ & $-1.6\times 10^{-3}$ \\
\hline \hline  
AV18+UX & VMC &   $-0.052$ &  $0.013$ & $0.017$ & $-1.6 \cdot 10^{-3}$\\
&GFMC &   $-0.047$ & $0.015$ & $0.019$ & $-2.0 \cdot 10^{-3}$ \\
\hline
AV18+IL7 & VMC &   $-0.053$ &  $0.013$ & $0.017$ & $-1.3 \cdot 10^{-3}$\\
&GFMC &   $-0.051$ & $0.016$ & $0.018$ & $-1.2 \cdot 10^{-3}$ \\
\hline
\end{tabular}
\caption{Pion-range matrix elements entering
$\overline{\delta^E_{\rm NS}}$,
the energy-dependent part of $\bar\delta_{\rm NS}$.}\label{TabII}
\end{table}

The matrix elements of the pion-range operators that contribute to $\delta_{\rm NS}^E$ are given in Table \ref{TabII}.  
The largest matrix element, $M^{E\pi}_{\text{GT}, p}$, enters $\delta_{\rm NS}^E$ multiplied by the prefactor $-\sqrt{2}\alpha R_A E_0\sim -2 \cdot 10^{-4}$, and thus it provides a  $\sim  10^{-5}$ correction, which is currently negligible.

\bibliography{delta}

\end{document}